\documentclass[acmlarge]{acmart}
\makeatletter
\newcommand{\confshort}{\acmConference@shortname}
\newcommand{\conffull}{\acmConference@name}
\newcommand{\confdate}{\acmConference@date}
\newcommand{\confloc}{\acmConference@venue}
\AtBeginDocument{
  \fancypagestyle{firstpagestyle}{
    \fancyhead{}%
    \fancyfoot[C]{}%
  }
  \fancyhf{}
  \fancyhead[LO]{\@headfootfont\shorttitle}%
  \fancyhead[RE]{\@headfootfont\@shortauthors}%
  \fancyhead[LE]{\@headfootfont\footnotesize \confshort, \confdate, \confloc}%
  \fancyhead[RO]{\@headfootfont\footnotesize \confshort, \confdate, \confloc}%
  \fancyfoot[C]{}%
}
\makeatother
\acmBooktitle{\conffull\@ (\confshort), \confdate, \confloc}

\AtBeginDocument{%
  }

\setcopyright{acmlicensed}
\copyrightyear{2026}
\acmYear{2026}
\acmDOI{10.1145/3805689.3812311}
\acmISBN{979-8-4007-2596-8/2026/06}
\acmConference[FAccT '26]{The 2026 ACM Conference on Fairness, Accountability, and Transparency}{June 25--28, 2026}{Montreal, Canada}

\begin{document}

\title[Taste for Privacy in Context]{Taste for Privacy: How Context, Identity, and Lived-Experience Shape Information Sharing Preferences}


\author{Juniper Lovato}
\email{jlovato@uvm.edu}
\orcid{0000-0002-1619-7552}
\affiliation{%
  \institution{University of Vermont}
  \city{Burlington}
  \country{USA}}
\affiliation{%
  \institution{Complexity Science Hub}
    \city{Vienna}
  \country{Austria}}

\author{Laurent H\'ebert-Dufresne}
\email{Laurent.Hebert-Dufresne@uvm.edu}
\orcid{0000-0002-0008-3673}
\affiliation{%
  \institution{University of Vermont}
  \city{Burlington}
  \country{USA}}
\affiliation{%
  \institution{Santa Fe Institute}
   \city{Santa Fe}
  \country{USA}}
\affiliation{%
  \institution{Complexity Science Hub}
    \city{Vienna}
  \country{Austria}}

\author{Mohsen Ghasemizade}
\email{Mohsen.Ghasemizade@uvm.edu}
\orcid{0009-0008-5758-1658}
\affiliation{%
  \institution{University of Vermont}
  \city{Burlington}
  \country{USA}}

\author{Jonathan St-Onge}
\email{Jonathan.St-Onge@uvm.edu}
\orcid{0000-0001-5369-4825}
\affiliation{%
  \institution{University of Vermont}
  \city{Burlington}
  \country{USA}}

\author{Peter S. Dodds}
\email{Peter.Dodds@uvm.edu}
\orcid{0000-0003-1973-8614}
\affiliation{%
  \institution{University of Vermont}
  \city{Burlington}
  \country{USA}}
\affiliation{%
  \institution{Santa Fe Institute}
   \city{Santa Fe}
  \country{USA}}
\affiliation{%
  \institution{Complexity Science Hub}
    \city{Vienna}
  \country{Austria}}

\author{Laura Bloomfield}
\email{Laura.Bloomfield@uvm.edu}
\orcid{0000-0003-0416-3440}
\affiliation{%
  \institution{University of Vermont}
  \city{Burlington}
  \country{USA}}

\author{Mikaela Irene Fudolig}
\email{Mikaela.Fudolig@uvm.edu}
\orcid{0000-0002-8960-9116}
\affiliation{%
  \institution{Adelaide University}
  \city{Adelaide}
  \country{Australia}}

\author{Matthew Price}
\email{Matthew.Price@uvm.edu}
\orcid{0000-0001-5637-9230}
\affiliation{%
  \institution{University of Vermont}
  \city{Burlington}
  \country{USA}}

\author{Christopher Danforth}
\email{Chris.Danforth@uvm.edu}
\orcid{0000-0002-9857-2845}
\affiliation{%
  \institution{University of Vermont}
  \city{Burlington}
  \country{USA}}

\renewcommand{\shortauthors}{Lovato et al.}

\begin{abstract}
Privacy preferences are not fixed individual traits, they depend on context and lived experiences. In this study, we analyze 2,912 survey responses from 782 college students collected over seven survey periods during 2023 and 2024.  We ask about their usage of social media, the security settings of their accounts, and measure their comfort in sharing personally identifiable information (PII) across 17 different institutional contexts.  Compared to past research, we observe a large shift towards private accounts, going from 1/3rd private in 2007 to 2/3rds in 2024, and find that participants' discomfort sharing PII with social media platforms strongly predicts their privacy settings.  Beyond social media, we identify a stable ranking of institutional trust, though some institutions, like the police, show high variability reflecting divergent lived experiences. Traditionally marginalized groups and participants having faced adverse childhood experiences show more discomfort with institutions of power, especially in areas where they face greater vulnerability. We argue for context-adaptive privacy settings that recognize institutional relationships and demographic vulnerabilities, moving beyond one-size-fits-all consent frameworks toward contextually appropriate data governance. 
\end{abstract}

\ccsdesc[500]{Security and privacy~Social aspects of security and privacy}
\ccsdesc[300]{Social and professional topics~User characteristics}
\ccsdesc[500]{General and reference~Surveys and overviews}
\ccsdesc[300]{Social and professional topics~Computing / technology policy}

\keywords{Privacy Preferences, Contextual Integrity, Personally Identifiable Information, Survey Study, Social Media Privacy, Institutional Trust}


\maketitle

\section{Introduction}
\label{secttion:introduction}

Privacy preferences are not a single fixed trait, an individual may be called a ``private person'' but this notion oversimplifies tastes for privacy. Rather, privacy preferences and tastes heavily depend on not only context but on who the person is and on their lived experiences within these contexts. A person may be private in one context and wish to be public in another based on their trust for that context and their prior experience sharing information within that sphere.    

Current privacy models assume stable individual preferences, but the taste for privacy varies across individuals and contexts and depends on the data recipient. Thus, our current notice-and-consent frameworks may be insufficient to capture the range of taste for privacy that individuals have in different contexts. Yet many online platforms still use one-size-fits-all privacy settings, ignoring this contextual variation.

Prior work by Lewis et al. \cite{lewis2008taste} measured the social media privacy settings and social and demographic patterns of college students in the Northeast US in 2007. They found that only 33.2\% maintained private Facebook profiles. They also found that women were significantly more likely than men to have private profiles, but found no significant association between privacy settings and race categories. In this new social media landscape, our study aims to measure if these patterns have changed and to examine how preferences vary across contexts, demographics, and lived experiences. 

In our study, we analyze 2,912 survey responses from 782 college students in the Northeast US collected over seven survey periods during 2023 and 2024. We measure participants' comfort in sharing personally identifiable information (PII) across 17 different institutional contexts and assess their social media behavior. In this study, we investigate the following research questions:

\begin{itemize}
    \item RQ1: Have the privacy preferences of college students changed since 2007? Further, how do social media privacy settings relate to comfort sharing personal data with platforms? And do usage patterns (frequency of use and number of platforms used) confound this relationship?
    \item RQ2: Do privacy preferences vary across institutional contexts? Further, is there a stable hierarchy of trust across institutions?
    \item RQ3: How do demographics and lived experiences shape patterns of comfort sharing PII with institutions?
\end{itemize}

In our results, we find that privacy preferences among college students have shifted since 2007, with 84\% now using private or mixed settings (62\% private, 22\% mixed) compared to just 33\% previously. We identify a stable hierarchy of institutional trust, with friends, medical professionals, and relatives consistently ranked as most trusted for most participants, while strangers and unfamiliar companies are least trusted. 

Social media platforms rank between unknown commercial entities and weak social ties, suggesting reluctant participation despite widespread use. Critically, discomfort sharing PII with social media platforms strongly predicts privacy settings, each step increase in discomfort raises the odds of more private settings by 37\%. We also find significant demographic variations reflecting lived experiences. Women show greater discomfort with acquaintances, medical professionals, and co-workers. People of color are more uncomfortable sharing with police, employers, and financial institutions. LGBTQ+ participants show heightened discomfort with neighbors, relatives, government, and police. Adverse childhood experiences systematically increase distrust of authority institutions, particularly relatives, financial institutions, police, medical professionals, employers, government, and neighbors.

This work is a measurement of contextual privacy preferences in college students, which may serve as evidence supporting contextual integrity preferences over privacy-as-trait models. We hope this work will demonstrate that one-size-fits-all privacy settings do not match the tastes for privacy of users who have different demographics and lived experiences. We argue that platforms should adopt context-adaptive privacy controls that allow users to specify what they would like to share with particular types of institutions and in what context. 

In section \ref{secttion:background} we explore the theoretical frameworks and empirical work that inform this study. In section \ref{secttion:methods} we outline our survey design, methods, and context, as well as our analytical methods. In section \ref{secttion:results} we outline our results by research question. In section \ref{secttion:discussion} we discuss the research implications, the limitations of the paper, and future work. In the Appendices (\ref{sec:appendix}) we provide detailed information on the survey questions, demographic breakdowns, descriptive statistics, and a full list of figures by demographic and institution type.

\section{Background}
\label{secttion:background}

\subsection{Theoretical Frameworks}

The dominant notice-and-consent model assumes individuals can make informed privacy decisions through the mechanism of a consent agreement. Nissenbaum and others critique this assumption, proposing a framework of contextual integrity \cite{nissenbaum2004privacy, barth2006privacy, nissenbaum2011contextual}. Privacy, she argues, must be evaluated within specific social contexts and their associated norms. Information flows that are appropriate in one setting (e.g., sharing health data with a doctor) may violate contextual privacy expectations in another (e.g., sharing the same data with an employer). Privacy preferences are context-dependent, and privacy judgments are shaped by social norms and power relations. 

Moreover, privacy preferences may not be a stable individual trait \cite{margulis2003status, bartol2025cross, smaldino2015evolution}. Martin explored the idea that online privacy norms and expectations shift by context \cite{martin2016understanding} and are seen as a social contract. He argues that instead of viewing privacy through notice-and-consent, it should be seen as a social contract that is a mutually beneficial agreement about how information flows. For Martin, privacy violations occur when information flows break the norms of the social contract. In this framework, information sharing is characterized by discriminate disclosure within contexts, where individuals maintain graded expectations about which information flows to which actors for what purposes.

Privacy is also not a simple risk-benefit tradeoff \cite{dinev2006extended} but a relational construct organized around boundaries between self and others \cite {altman1975environment, bucher2019algorithmic, masur2018situational, kassam2023patient}. Altman conceptualized privacy as a dynamic process of regulating boundaries. This was a process where individuals selectively control access to the self across different relationships and contexts. People navigate social and normative preferences through comfort zones (from close friends to strangers), where different norms govern each circle. Privacy scholars have explored this relational framework, showing that privacy boundaries shift based on relationship type \cite{petronio2002boundaries}, serve distinct psychological functions across social contexts \cite{pedersen1997psychological}, and require continuous calibration depending on situational factors \cite{masur2018situational}, and trust in different organizations varies based on benevolence, integrity, and propensity to trust  \cite{mayer1995integrative}.

Privacy preferences are not always reflected in privacy behavior, as seen in the privacy paradox literature \cite{acquisti2005privacy, kokolakis2017privacy, norberg2007privacy, hargittai2016can}. When faced with the social optimization problem of trading long-term privacy for near-term benefits, consumers often forgo privacy even if they have a strong taste for privacy. This challenges the idea that consumers are rational actors in privacy decision-making. In Acquisti et al. \cite{acquisti2005privacy}, a survey of 119 respondents was conducted to assess their privacy concerns and privacy behavior. The privacy paradox states that even when consumers have strong privacy concerns, their behavior contradicts those concerns. Draper et al. explain the privacy paradox \cite{draper2019corporate} as a rational, reluctant participation or digital resignation, arising from institutional power asymmetries in which consumers want to control the flow of their information but feel unable to do so. This resignation is not just a natural response to surveillance, but companies actively cultivate helplessness through intentional obfuscation (via placation, diversion, jargon, and misnaming). 

We operationalize these concepts by measuring how the same individuals vary their comfort sharing PII across 17 institutions. Importantly, privacy preferences are not fixed: different demographic groups may position institutions differently based on lived experiences. Our analysis examines both the shared structure of privacy preferences across college students and the demographic variations that reflect how identity shapes institutional proximity in the privacy landscape.

\subsection{Empirical Studies of Privacy Preferences}

Several studies have explored social media privacy preferences of college students \cite{gross2005information, lewis2008taste}. Our work builds on and extends prior research that assesses privacy settings and demographic patterns in social media use of college students. \cite{lewis2008taste} Lewis et al. analyzed privacy behavior among 1,710 students at a private university in the Northeast of the United States, finding that 33.2\% maintained private Facebook profiles by summer 2007, while the others had public profiles. The study found that students with more friends and roommates who had private profiles were significantly more likely to maintain their own private profiles. These findings support the premise that privacy preferences are shaped by lived experience. Additionally, they found that women were significantly more likely than men to have private profiles, though no significant association was found between privacy settings and race categories. 

In a study conducted by Shane-Simpson et al. the authors investigate why 664 college students use different social media platforms \cite{shane2018college}, they explored the reasons behind using three popular social media platforms, Instagram, Facebook, and Twitter (now known as X), based on personal characteristics such as gender, age, features of specific sites, and privacy concerns that predicted social media preferences. They found that younger men use Twitter more frequently than other groups, and Twitter users tend to have public profiles rather than private ones. Privacy concerns were greater among Facebook users than Twitter users, suggesting that students may use Facebook partly because it offers more privacy controls. The privacy settings, communication modalities, and types of social connections available on each site may appeal to different types of individuals. 

Privacy preferences also relate to identity \cite{schlesinger2017intersectional}, past relationships with surveillance, power, and vulnerability in society. Historically marginalized groups will have a different perception of risk when sharing information with certain organizations. Historically and currently marginalized groups have been disproportionately targeted by surveillance \cite{browne2015dark, gilliom2001overseers, benjamin2023race}, and this has created institutional distrust among marginalized groups toward surveillance in general and towards specific institutions and organizations \cite{eubanks2018automating}.

Research on gender differences in privacy preferences shows consistent patterns across contexts. Pederson's Privacy study surveyed 260 college students and found that women showed significantly higher preferences for intimacy with family and friends, while men preferred isolation, though no gender differences emerged for reserve, solitude, or anonymity \cite{pedersen1987sex}. These preference differences translate into actual privacy behaviors on social media platforms. Stutzman et al.'s 2008 study of UNC Chapel Hill undergraduates found that males were only 59\% as likely as females to maintain friends-only Facebook profiles, even though those with private profiles had significantly more friends overall \cite{stutzman2010friends}.

However, privacy-setting behavior varies not only by gender but also by age and familiarity with platform policies. Kanampiu et al.'s exploratory study of 153 Facebook users across age groups found that younger users (18–24) were more proficient at privacy-related tasks than older users (55+), and that women were more familiar with Facebook privacy policies than men \cite{kanampiu2018privacy}. Notably, education level did not significantly impact privacy-setting behavior. 

People who have different lived experiences and identities interact with technology and privacy in meaningfully different ways. Vulnerability to privacy violations through synthetic media varies by demographics, with individuals showing greater accuracy detecting manipulated content that matches their own demographic attributes \cite{lovato2024diverse}. Beyond platform settings, preferences for data sharing vary by professional identity and practice, for example, artists show differential comfort with AI systems using their work depending on their artistic medium, demographics, and experience using the technology \cite{lovato2024foregrounding}.

In our study, we build upon Lewis et al.'s (2008) finding \cite{lewis2008taste} and assess to what extent college students' privacy preferences have changed over time. We measure privacy preferences across 17 institution types, allowing us to examine how the same individuals vary their comfort sharing personally identifiable information by institutional context. Finally, we analyze demographic variation not as an isolated effect but as patterns that reveal how lived experiences shape privacy preferences.

\section{Methods}
\label{secttion:methods}

\subsection{Study Design and Context}


This study was a sub-study embedded in the Lived Experiences Measured Using Rings Study (LEMURS) at the University of Vermont \cite{price2023large}. The broader study examines the behaviors and well-being of young adults, focusing on undergraduate students in the Northeast United States. The goal of the broader study is to understand the variation in behaviors, social connections, and well-being of college students to inform strategic interventions that promote long-term health and longevity. The broader study investigates how students' experiences, mentorship, and social dynamics shape their academic and personal development. The study employs a series of surveys capturing demographic information (baseline surveys captured at the beginning of a new cohort enrollment period), as well as behavioral, psychological, and health-related measures. In seven of these surveys, we assess students' attitudes toward privacy and information-sharing preferences of personally identifiable information (PII), providing insights into data sharing preferences, privacy behaviors, and concerns. Surveys are administered on a weekly basis. Questions related to privacy preferences are only asked during baseline, mid-term evaluations, and semester-end surveys. Surveys for this paper were conducted during 2023–2024. This study protocol was reviewed and approved by the author's university Institutional Review Board (protocol number 00002126).

Participant eligibility for the study required incoming participants to be first-year undergraduates who were 18 years old or older and had a compatible mobile device. Recruitment was conducted online and on campus. Advertisements promoting the study were distributed across campus. Interested students accessed a QR code linking to a brief online eligibility survey. Eligible participants received an online informed consent form, a brief explanatory video covering study procedures, risks, and benefits, and a comprehension quiz to ensure informed consent. Following enrollment, participants attended an in-person orientation and received an Oura Ring and a mobile app through which the study's instructions and digital activities were delivered. Participants were compensated for their participation and successful completion of the study activities.

\subsection{Survey}

Seven total surveys from the broader project were used in this study, namely the ones that ask questions about taste for privacy. These surveys were collected over seven survey periods: 1/2023 - baseline (406 unique responses), 3/2023 (328 unique responses), 4/2023 (291 unique responses), 7/2024 - demographics (580 unique responses), 1/2024 (455 unique responses), 4/2024 (456 unique responses), 5/2024 (396 unique responses). There were a total of 782 unique survey respondents with a total of 2,912 unique responses over the 7 periods. 

The survey items related to taste for privacy on social media comprised 3 questions that explored students' social media settings, which platforms they use, and the frequency of use. The exact survey questions and response options are shown in Table \ref{tab:tablesurveyt4p}.

\begin{table*}
    \centering
    \begin{tabular}{p{0.45\textwidth}p{0.45\textwidth}}
    \toprule
       Question  & Response Options  \\
       \midrule
       Are your social media profiles typically public or private?  & Public, Mixed, Private.   \\
       
        Which platforms do you use? &  (check all that apply) Twitter, Instagram, Facebook, TikTok, Other. \\
        
        In a typical month, how often do you use social media? &  More than once a day, Once a day, A few times a week, Once a week, A few times a month, Once a Month, I use social media but less than once a month, I do not use social media. \\
    \bottomrule
    \end{tabular}
    \caption{Taste for privacy survey questions}
    \label{tab:tablesurveyt4p}
\end{table*}

After the social media questions, we expand to more contexts and ask the participants ``How comfortable are you with sharing your personally identifiable data with a'' and we list 17 institutions as response options ``Government official, Police Officer, Friend, Relative, Employer, Medical Professional, Financial Institution, Neighbor, Acquaintance, Co-worker, School, Researcher, Social media platform, Non-profit, Company where you are a customer, Company where you are not a customer, and Stranger.'' Participants rated each institution on a 7-point Likert scale, where 1 = Very comfortable, 2 = Comfortable, 3 = Slightly Comfortable, 4 = Neutral, 5 = Slightly Uncomfortable, 6 = Uncomfortable, 7 = Very uncomfortable. These 17 institutions were chosen to reflect a range of potential relationships from close personal (Friend, Relative), professional and formal (Employer, School, Researcher, Financial institution, Medical professional), civic (Government official, Police officer, Nonprofit), commercial (Platform, Company (customer), Company (not customer)), and weak social ties (Neighbor, Acquaintance, Coworker, Stranger). 

Participants are also asked about their demographics and lived experiences in the two baseline surveys. Participants were asked to answer questions about standard demographics (gender, race/ethnicity, and sexual orientation) in the demographics questions; note that multiple responses per category are possible if more than one category is true for an individual. Race categories were coarse-grained into White, Mixed, and Person of Color due to small sample sizes across the available categories. All demographic variables were one-hot encoded to convert them into numeric form for analysis. Participants who selected gender categories other than male or female were grouped into a single non-binary category due to small sample sizes that would otherwise preclude meaningful statistical analysis. This aggregation was necessary because individual demographic categories had insufficient sample sizes for separate analysis, and some participants' selections varied across survey rounds. A breakdown of all demographics can be seen in Appendix \ref{secttion:demographics}. 

Questions were also asked about contextual factors (First-generation status, therapy participation) and adverse childhood experiences (Adverse Childhood Experiences Scale, ACE) \cite{Karatekin2023}, which assesses childhood exposure to household dysfunction, abuse, and neglect. A full list of ACE questions can be found in the appendix \ref{secttion:completesurvey}.

\subsection{Analytical methods}

In Research Question 1 (RQ1), we explore privacy settings and platform discomfort using an ordinal logistic regression (privacy settings as outcome: Public, Mixed, Private) vs. comfort sharing PII with social media platforms. From the ordinal logistic regression, we calculate the odds ratios \cite{agresti2012categorical} for moving to a more private setting per unit increase in discomfort sharing PII with social media companies. To explore potential confounding variables, we extend the model to include frequency of social media usage in a typical month and the number of social media platforms used, alongside discomfort sharing PII with social media platforms. 

In Research Question 2 (RQ2), we rank institutions by participants' comfort sharing PII. Rankings are based on participant ordering of Likert scores, with institutions sorted by mean rank. We display these rankings using violin plots showing the distribution and quartiles of ranks across participants.

In Research Question 3 (RQ3), we first conduct difference-in-mean tests with bootstrap confidence intervals (10,000 bootstrap samples) for pairs of demographic groups (e.g., Women vs. Men). We compute the mean difference in discomfort for all 17 institutions and bootstrap 95\% CIs for each demographic pair. We then conduct a bootstrap comparison (10,000 bootstrap samples) of institution rankings between demographic groups. For each bootstrap iteration, we independently resample each demographic group, calculate the mean discomfort per institution within each group, and rank institutions from most to least uncomfortable (rank 1 = highest mean discomfort). We compute the difference in rank positions between groups for each institution (Group 1 rank - Group 2 rank), where positive values indicate the institution ranks as more uncomfortable for Group 1. We calculate 95\% confidence intervals for these rank differences across bootstrap iterations.

To explore adverse childhood experiences (ACEs), which vary in number and type per person, we conduct a dose-response analysis \cite{agresti2010analysis}. We fit an ordinal logistic regression model per institution with discomfort as the outcome and ACEs dose as the predictor. The resulting coefficient represents the change in log-odds of being in a higher discomfort category per additional ACE. We visualize these effects using a forest plot showing coefficients and 95\% confidence intervals across all institutions.

\section{Results}
\label{secttion:results}

We studied 2,912 survey responses collected during seven survey periods from 782 college students over a two-year period. The multiple answers from single participants are not used as a longitudinal study, but to increase robustness by averaging answers. 

Lewis et al. found in 2008 that a third of online social network accounts belonging to college students had a private profile, and two-thirds had a public profile \cite{lewis2008taste}. In our surveys, we find that only 16\% have a public profile. Other accounts are mostly private (62\%) or with mixed settings (22\%). We also study how those changes reflect a discomfort with social media platforms themselves, as well as with third parties that could access the data. Importantly, we find that this discomfort reflects a general hierarchy of (dis)trust towards different institutions, but can also vary greatly by demographics and lived experiences. 

\subsection{RQ1: Privacy Settings \& Platform Discomfort}
\label{secttion:resultsrq1} 

\begin{figure}[h]
    \centering
    \includegraphics[width=0.7\columnwidth]{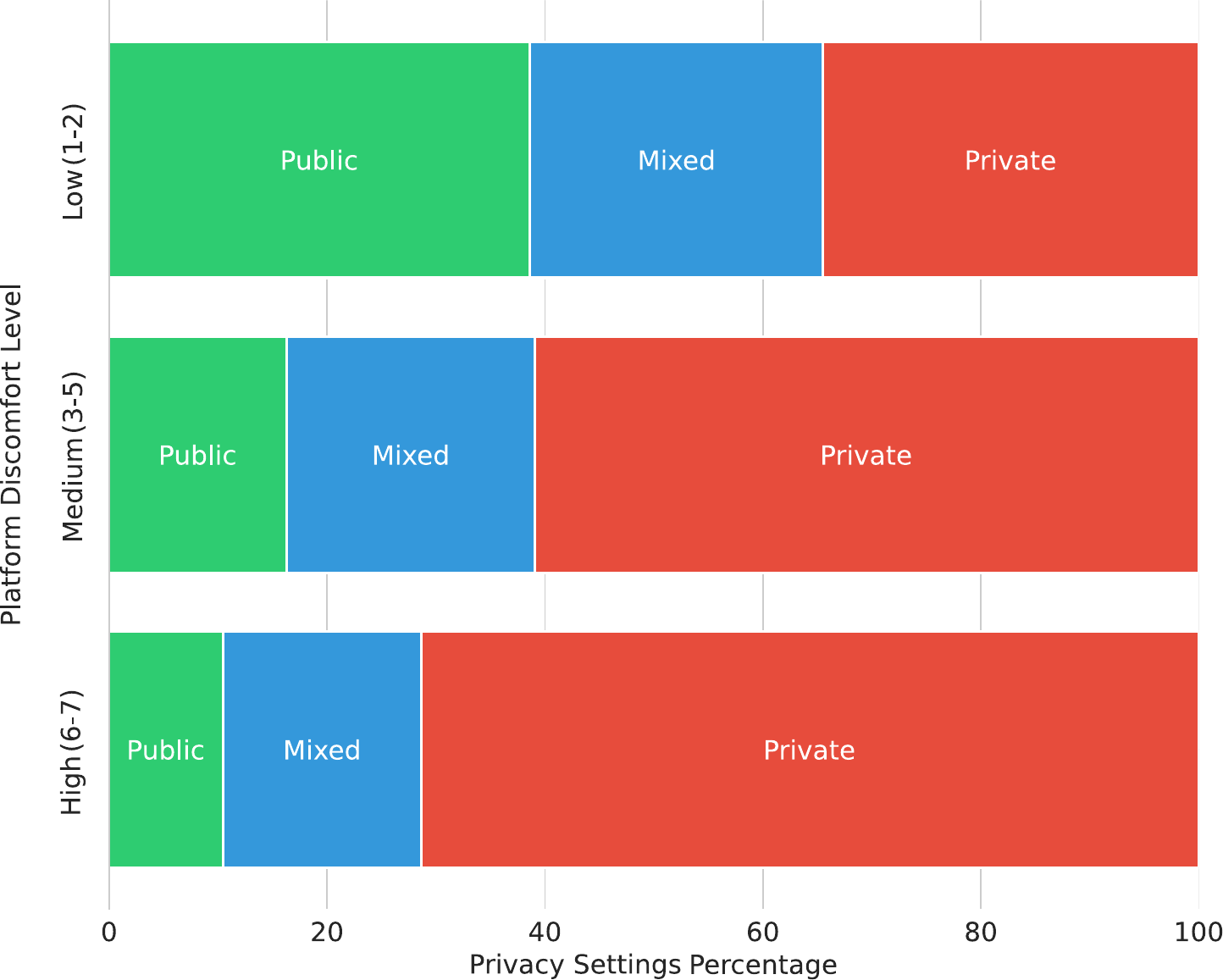}
    \caption{Stacked bar chart showing the proportion of participants using public (green), mixed (blue), or private (red) privacy settings on social media, stratified by comfort sharing PII with social media platforms (7-point Likert scale collapsed into three categories).}
    \Description{Stacked bar chart showing three horizontal bars representing low, medium, and high platform discomfort levels. Each bar is divided into three colored segments: green for public settings, blue for mixed settings, and red for private settings. The low discomfort bar shows roughly equal proportions of all three settings. The medium discomfort bar shows decreased public use. The high discomfort bar shows minimal public use. A clear trend shows that increased discomfort correlates with more private settings.}
    \label{fig:privacysettingbylikertgroups}
\end{figure}

The proportion of private social media accounts among college students was estimated to be 33\% in 2008 \cite{lewis2008taste}. This figure appears to have increased to 62\%, or up to 84\% if we include accounts described as using mixed settings. Essentially representing a complete flip in the proportion of college students with a ``taste for privacy.'' 

To investigate what features of participants could help predict their privacy settings, we asked them how much they use social media (8 point scale ranging from ``I do not use` to ``more than once a day''), how many platforms they use (Twitter/X, Instagram, Facebook, TikTok, or other), and how comfortable they are sharing data with social media platforms (7 point Likert scale). We used an ordered model (or proportional odds ordinal logistic regression) to parse out the effects of these variables. In this regression, an odds ratio (OR) smaller than 1 represents a negative association between the variables. This analysis shows that, controlling for comfort level, both variables related to social media usage independently predict less private settings. We find a stronger effect for the number of platforms used (OR=0.82, $p$-value $<$ 0.001) than for usage frequency (OR=0.90, $p$-value $<$ 0.020). This means that participants who use social media more are more likely to have public accounts. This result is perhaps intuitive, but unfortunate for the overall privacy of these participants. 

Looking at the (dis)comfort of participants with social media platforms, we performed a similar ordered model and found that students who are less comfortable sharing data are significantly more likely to use private (vs mixed) or mixed (vs public) privacy settings. A greater discomfort sharing with the platform strongly predicts more private social-media settings: OR=1.37 per Likert step ($p$-value $<$ 0.001, 95\% CI 1.30–1.43).
For each point of discomfort with the platform (e.g., from 4 ``neutral'' to 5 ``slightly uncomfortable''), this implies that the odds of using more private settings increase by ~37\%, implying ~3.5× higher odds for a 4-step increase and ~6–7× for a 6-step increase. We thus find that students who feel less comfortable sharing with the social media platforms are much more likely to choose private global social settings; conversely, the chance of public settings falls as discomfort rises.

Importantly, when controlling for social media usage and number of platforms used, greater discomfort with the platform still strongly predicts more private global settings (OR=1.34 per Likert step, $p$-value $<$ 0.001, 95\% CI 1.27–1.40).
As a summary of this association, the overall proportions of privacy settings per discomfort level are presented in Fig.~\ref{fig:privacysettingbylikertgroups}.

\subsection{RQ2: Institutional Trust Hierarchy}
\label{secttion:resultsrq2} 

\begin{figure*}
  \centering  
\includegraphics[width=.55\linewidth]{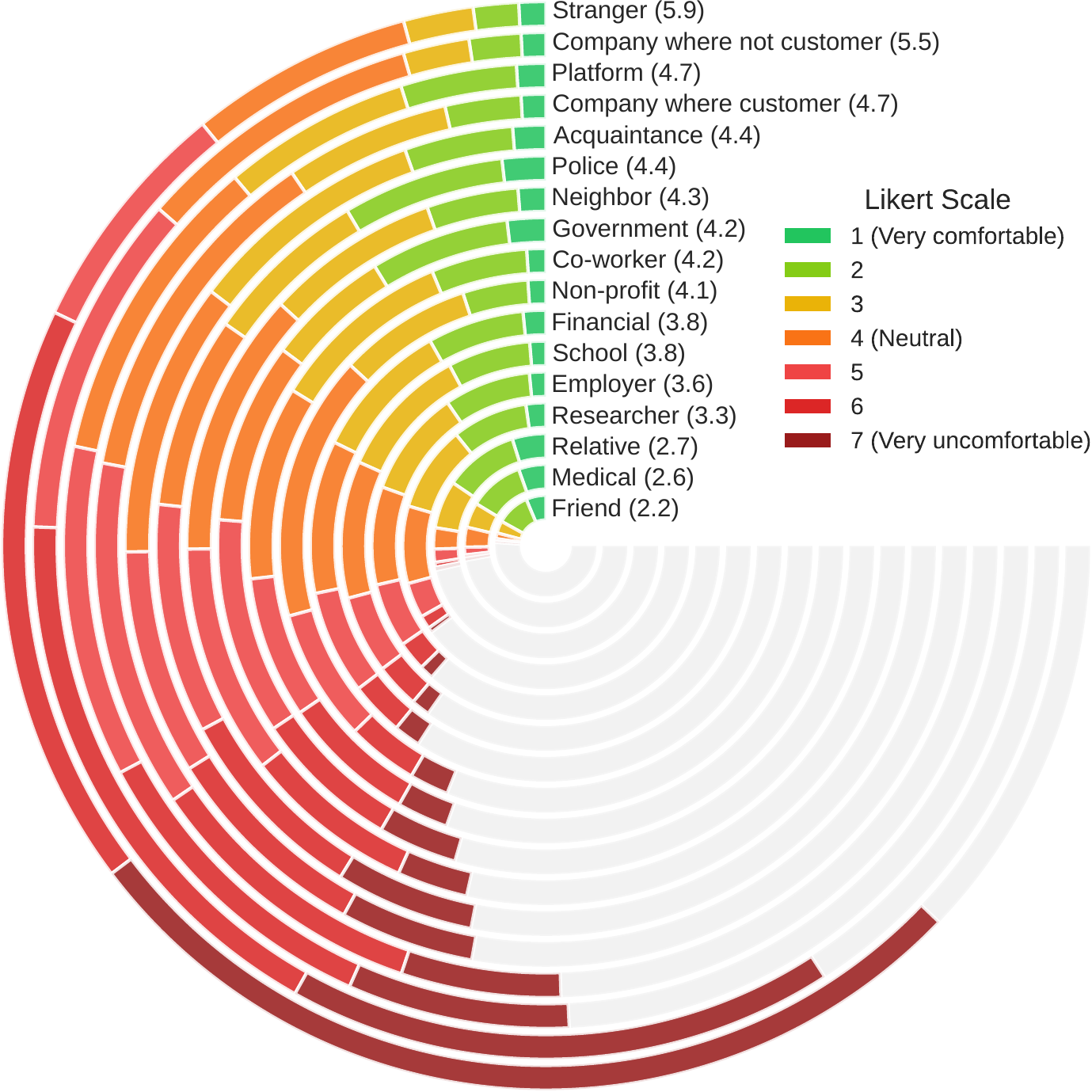}
  \caption{Comfort Sharing PII by Institution. Arc length represents the mean discomfort level, ranging from 1 when very comfortable (green) to 7 when very uncomfortable (red). Colored segments show the full breakdown of Likert responses.}
  \Description{Circular stacked bar chart showing comfort levels for sharing personal information with 17 institutions. Each institution is represented by a horizontal bar bent into an arc, divided into colored segments for the 7-point Likert scale. Green segments indicate comfort, yellow/orange neutral responses, and red segments discomfort. Strangers show the longest red segments and highest mean discomfort (5.9). Friends show the longest green segments and lowest mean discomfort (2.2). The pattern reveals concentric trust circles with close relationships (friends, medical, relatives) toward the center and distant entities (strangers, companies) at the outer edge.}
  \label{fig:rankingcircle}
\end{figure*}

\begin{figure*}
    \centering
    \includegraphics[width=.55\linewidth]{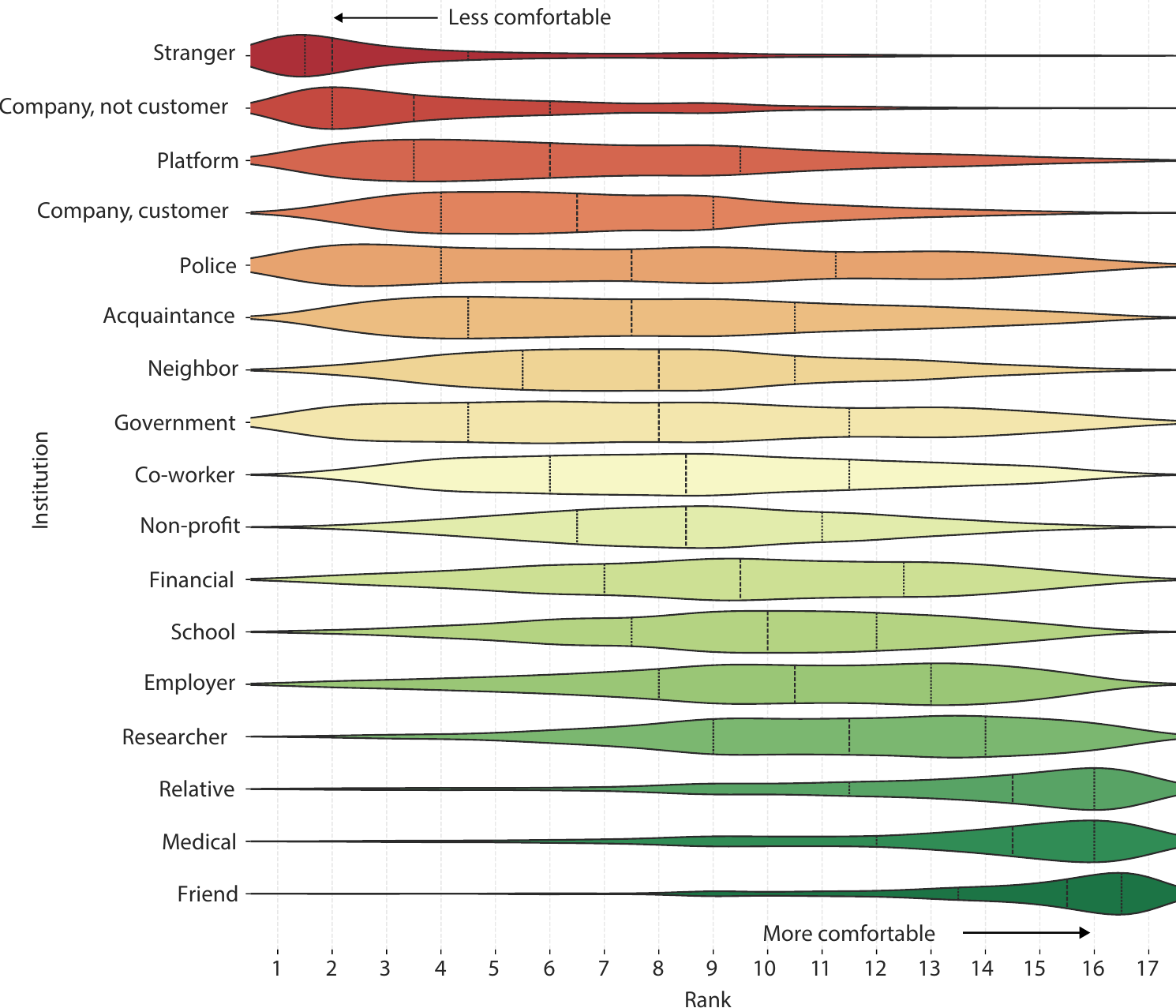}
    \caption{Participant rankings of institutions by discomfort sharing PII. Each violin shows the distribution of individual rankings (1 = least comfortable, 17 = most comfortable). Dashed lines indicate quartiles; color represents mean rank. Width indicates frequency of rank assignments.}
    \Description{Violin plot showing how participants ranked 17 institutions by discomfort sharing personally identifiable information. Strangers, non-customer companies, and social media platforms cluster at high discomfort ranks (1-6) with narrow red/orange distributions. Friends, medical providers, and relatives cluster at low discomfort ranks (14-17) with narrow green distributions. Middle institutions show wider, more varied distributions, indicating less consensus among participants.}
    \label{fig:ranksofinstitutionsbootstrapped}
\end{figure*}

Other than social media platforms, the survey asked participants about 16 other third-party users or institutions that could potentially access their data. The breakdown of comfort sharing data (7-point Likert scale) over all of the 17 institutions and the entire population of participants is presented in Fig.~\ref{fig:rankingcircle}. Not surprisingly, we find that participants are most comfortable sharing data with friends and least comfortable sharing data with strangers. Some of the more interesting results are found in-between these extremes, with medical and research institutions being the second and fourth most comfortable institutions (although, as we will discuss, this can be an artifact of the studied population). Interestingly, for-profit companies where the participant is a customer receive very similar answers as social media platforms themselves; potentially clashing with the common representation of social media as a business where users are not customers but the product.

The values of the Likert scale can be quite noisy as some participants are intrinsically more or less comfortable sharing data than others (the standard deviation for the average Likert per institution ranges from 1.168 to 1.714). We therefore also look at the stability of overall ranking of institutions in Fig.~\ref{fig:ranksofinstitutionsbootstrapped}. We find that three institutions are consistently marked as comfortable: friends, medical professionals, and relatives.
Next, is a cluster of institutions with a clear average ordering and generally high comfort level but less consistent absolute ranks; in order of comfort rank: researchers, employers, schools, financial institutions, and non-profit institutions.
Next, we find a cluster of interchangeable institutions; in order of average comfort rank: co-workers, government, neighbors, acquaintances, police, company where the participant is a customer, and the social media platforms.
Finally, two institutions are consistently marked as uncomfortable: strangers and companies where the participant is not a customer.
The uncertainty in ranking within the two middle clusters (and especially for the police) reflects differences across demographics that we study in the next section.

The clear takeaway from these results is that privacy preferences are not a blanket, static, or average property of individuals themselves. They vary greatly across relationship type and on a complicated continuum.
On average, we find that social distance between the participant and the institution decreases comfort level. 
Discomfort grows when going from friend to relative to coworker to acquaintance and finally to stranger.
Taste for privacy is therefore not an individual trait, but a property of the relationship between individuals or between an individual and an institution.
Our results highlight a generally stable ranking of these relationships, but with variance within some particular clusters that we now investigate.

\subsection{RQ3: Demographic Variations}
\label{secttion:resultsrq3} 

\begin{figure*}[h]
    \centering
    \includegraphics[width=1\linewidth]{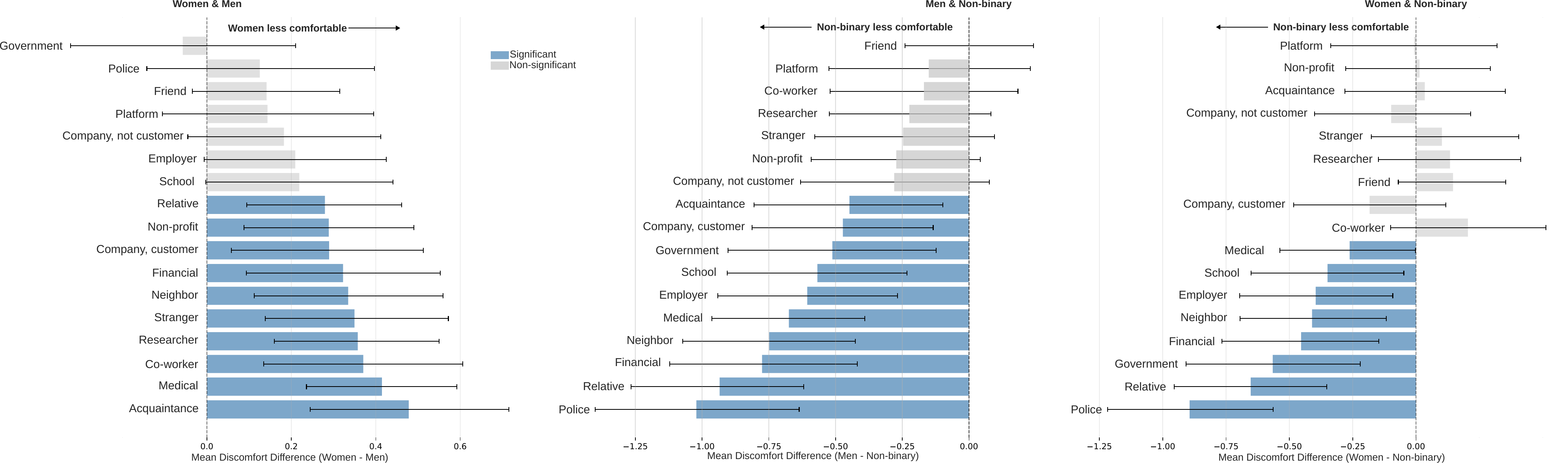}
    \caption{Differences in mean discomfort sharing personally identifiable information across demographic groups. Forest plots showing bootstrapped mean differences in discomfort (with 95\% confidence intervals) for 17 institutional contexts across three demographic comparisons: Women vs. Men (left), Men vs. Non-binary (center), and Women vs. Non-binary (right). Positive values indicate the first group is less comfortable (more discomfort); negative values indicate the second group is less comfortable. Blue bars indicate significant differences (95\% CI excludes zero); gray bars indicate non-significant differences.}
    \Description{Three side-by-side forest plots comparing discomfort levels across demographic groups. Left panel compares women versus men, showing women significantly more uncomfortable with 10 of 17 institutions. Center panel compares men versus non-binary individuals, showing non-binary individuals significantly more uncomfortable with 10 institutions. Right panel compares women versus non-binary individuals, showing non-binary individuals significantly more uncomfortable with 8 institutions. All significant differences are marked in blue with confidence intervals excluding zero; non-significant differences appear in gray.}
    \label{fig:wgendermeandiffbootstrapped}
\end{figure*}

The surveys collected data about the demographic and lived experiences of participants, which can help explain the variability observed in comfort rankings of institutions.
To do so, we first calculate distributions of differences in (dis)comfort level between pairs of demographics. Many of these differences are highlighted in our Appendix. 
The main result from these analyses is shown in Fig.~\ref{fig:wgendermeandiffbootstrapped} and reflects gender-based differences in data sharing comfort. 

Recall that in 2008, Stutzman et al. found that women were almost twice as likely than men to have private social profiles \cite{stutzman2010friends}. We find a somewhat similar but more nuanced result. Women are significantly less comfortable sharing data with some close social connections, such as acquaintances, co-workers, neighbors, and relatives. Interestingly, the second biggest difference is that women are also less comfortable sharing data with medical professionals. Otherwise, bigger institutions such as social media platforms, police, and government show no significant differences between men and women.

Our survey allowed participants to select other gender classifications beyond the binary distinction of men and women. For statistical reasons, because of small group sizes, and because participants' answers could change between survey rounds, we unfortunately group those participants under a very broad non-binary demographic. When comparing the non-binary demographic to either men or women, we find them generally less comfortable sharing data with both close and distant social connections: neighbors, relatives, government, and police all showing statistically significant differences. In fact, the discomfort of the non-binary demographic in sharing data with police is the single biggest difference observed across all demographics.

Importantly, there are no institutions or relationships in which men are less comfortable sharing data than any other demographic. Likewise, there are no institutions or relationships in which non-binary participants were more comfortable sharing data than any other demographic.

Looking at other demographics, we find that people of color were more uncomfortable than white people sharing data with police (average difference 0.65, $p$-value $<$ 0.001), government (average difference 0.52, $p$-value $<$ 0.01), and strangers (average difference 0.38, $p$-value $<$ 0.05). Across sexual orientation, the biggest differences observed show homosexual participants being less comfortable than heterosexual participants in sharing data with medical professionals (average difference 0.71) or with their employer (average difference 0.58). First-generation students are more uncomfortable sharing data with financial institutions (0.42), their schools (0.38), and researchers (0.35). Results for all demographics are shown in the Appendix \ref{sec:appendix}.

In terms of lived experiences, we first find that students in therapy are more comfortable sharing data with medical professionals (average difference 0.48, $p$-value $<$ 0.001) and with researchers (average difference 0.32, $p$-value $<$ 0.05). This result suggests that an increased familiarity with health and research contexts reduces discomfort.

\begin{figure*}[h]
    \centering
    \includegraphics[width=.58\linewidth]{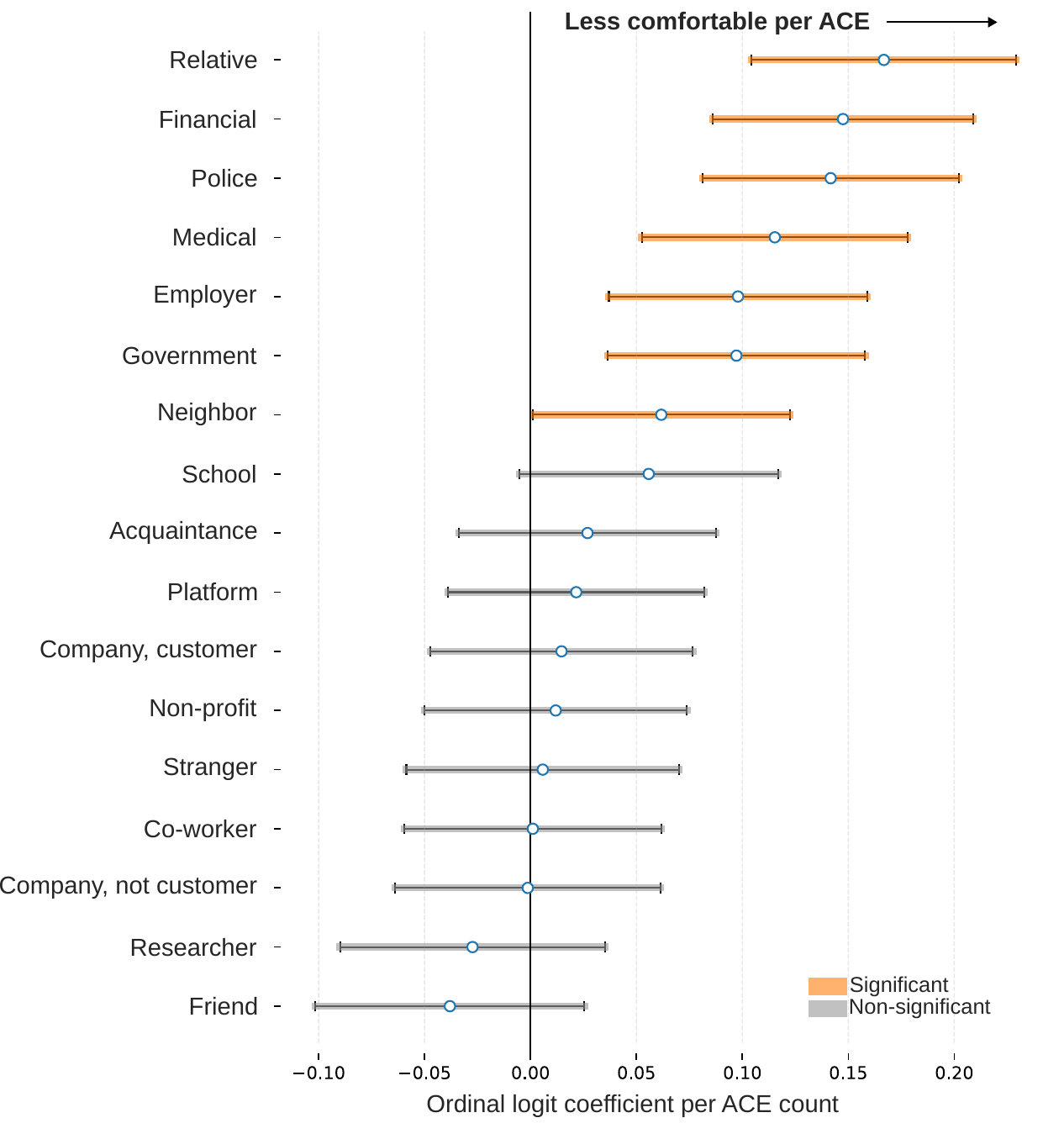}
    \caption{Dose-response relationship between ACE count and institutional discomfort. Forest plot showing ordinal logistic regression coefficients (change in log-odds per additional ACE) with 95\% bootstrap confidence intervals. Orange: significant associations; gray: non-significant. Higher ACE exposure is associated with greater discomfort sharing PII with relatives, financial institutions, police, medical providers, employers, and government agencies.}
    \Description{Forest plot showing the relationship between adverse childhood experiences and discomfort sharing personal information across 17 institutional contexts. Seven institutions show significant positive associations (orange), with relatives, financial institutions, and police showing the strongest effects. Ten institutions show non-significant associations (gray), clustered near zero.}
    \label{fig:acesdosebootstrapped}
\end{figure*}

We also asked participants about their Adverse Childhood Experiences (ACEs). We produced a composite score capturing the different types of ACEs faced by a participant and used an ordinal logit regression to estimate how ACEs affect their comfort sharing data. These results are shown in Fig.~\ref{fig:acesdosebootstrapped}.
We find that each ACE increases discomfort level in sharing data with specific institutions, in order of effect size: relatives, financial institutions, police, medical professionals, employers, government, and neighbors.

ACEs appear to increase distrust toward institutions and relationships that reflect a position of power or authority. In that regard, the category of relatives would be interesting to break down in future work. The effects measured are concentrated in contexts where vulnerability is heightened (e.g., relatives, financial institutions, police, medical professionals, employers, government, and neighbors). Conversely, close relationships (e.g., friends) and distant neutral ones (e.g., strangers) are unaffected.

\section{Discussion}
\label{secttion:discussion} 

Our study demonstrates that the same individuals vary systematically in their comfort sharing PII across different institutional contexts, and that this variation is not random but patterned by the type of institution, the individual's demographics, and their lived experiences. This challenges the assumptions underlying current notice-and-consent frameworks, which treat privacy preferences as stable individual characteristics that can be captured through blanket agreements across many contexts. 

Our study demonstrates three primary findings. First, that privacy behavior among college students has shifted dramatically since Lewis et al.'s 2008 study. We find that 62\% now maintain fully private social media profiles (compared to 33\% in 2007), with an additional 22\% using mixed settings. This shift correlates strongly with platform discomfort, where each step increase in discomfort sharing PII with social media platforms raises the odds of more private settings by 37\%.

Even though students have increased use of private social media settings and demonstrate growing discomfort sharing data with social media platforms, they still use social media daily. Students who use more platforms and use them more frequently are significantly more likely to maintain public settings, suggesting a tension between privacy preferences and social demands. This pattern supports Draper and Turow's concept of digital resignation \cite{draper2019corporate}. The relationship between platform discomfort and private settings suggests these are not hypocritical users who claim to care about privacy but act otherwise, rather, they are individuals managing an impossible social optimization problem that presents a tradeoff between long-term privacy concerns and immediate social connection.

Second, we identify a stable hierarchy of institutional trust across 17 institutions. Three institutions consistently rank as most trusted: friends, medical professionals, and relatives. A middle tier includes researchers, employers, schools, financial institutions, and non-profit organizations. A lower tier encompasses co-workers, government, neighbors, acquaintances, police, and customer-relationship companies. At the bottom sit strangers and companies where no relationship exists. Social media platforms, interestingly, are positioned between weak social ties and unfamiliar companies. 

Third, demographic patterns show that marginalized groups experience their discomfort most acutely in contexts where they face heightened vulnerability. Women show greater discomfort with weak social ties and medical professionals. People of color are more uncomfortable sharing with police, employers, and financial institutions. LGBTQ+ participants show heightened discomfort with neighbors, relatives, government, and police. The adverse childhood experiences (ACEs) findings demonstrate how trauma systematically reshapes institutional trust. Our dose-response analysis reveals that each additional ACE increases discomfort with institutions of authority and power: relatives, financial institutions, police, medical professionals, employers, government, and neighbors. The impact is concentrated in contexts of power asymmetries and potential vulnerability, suggesting that early experiences with institutional failure or harm create lasting wariness toward similar power structures.

These patterns suggest that privacy preferences reflect lived experiences of institutional relationships rather than stable individual characteristics, challenging one-size-fits-all consent frameworks. We hope this work informs the design of new privacy settings and data governance models that enable context-adaptive privacy, recognizing both institutional trust hierarchies and demographic vulnerabilities. Such systems should provide meaningful defaults that reflect our measured trust hierarchies while allowing personalization. Privacy settings that adapt by context \cite{6263765, 10.1145/3613904.3642500, shaffer2021applying, alamari2025adaptive} could enable individuals to express their tastes for privacy by context and data recipient, for example, on social media platforms, a context-adaptive privacy setting would allow users to specify what kinds of advertisers or other users are allowed to access their data. This contextual variation also extends to networked privacy, where individual preferences cannot be exercised independently because information flows through social connections \cite{lovato2022limits}.

In addition to platform settings, this may impact policy on privacy, as the adoption of more context-aware consent legal requirements may better reflect varying tastes for privacy. These may involve implementing appropriate standards and thresholds across contexts making it easier to opt out of high-discomfort contexts. In addition, because users do have a taste for privacy in different context they should be aware of which contexts their data flow. Policy should extend to increasing transparency about data flows to institution types and show users not just who is accessing their data but what kinds of data they are accessing. 

\section{Limitations and Future Work}

There are several limitations of this work. The first is that participants who opt into this study may have a lower taste for privacy since they are more willing to share their PII with researchers, which may represent a Self-Selection Bias. This study population may not generalize to a broader representative sample, as this is a study conducted with college students at one Northeast U.S. University on a largely white student body.

In future work, we would like to expand this study to a more representative general population sample. An extension could explore not just what institutions participants feel comfortable sharing PII with, but what kinds of data types they feel comfortable sharing by institution. We would like to examine the longitudinal effects of privacy preferences, including the effects of being involved as a participant in such a study. Finally, we also aim to conduct long-form interviews with participants of the present study to gain a deeper insight into their privacy preferences and how their lived experience shapes their tastes for privacy, and how they reason about privacy decisions in different contexts. These long-form interviews will help us to understand more deeply why certain institutions elicit varied preferences, and what makes them trustworthy or not trustworthy to this population.

\section*{Generative AI Usage Statement}
Writing for this manuscript was performed by the authors. Some authors acknowledge use of LLM powered technology (Writefull for grammar and spell checking) for refinement of their original text. 

\begin{acks}
The author would like to thank the reviewers for their feedback on this project. 
We would also like to thank the LEMURS team for their work assisting with this broader project. We would also like to thank the study participants for their time. Our team acknowledges support from the MassMutual Center of Excellence in Complex Systems and Data Science (Grant \# FP2860), Alfred P. Sloan Foundation (Grant \#G-2024-22498), and the National Science Foundation (Grant \#2242829) 
\end{acks}

\bibliographystyle{ACM-Reference-Format}
\bibliography{bib}

\appendix
\label{sec:appendix}

\section{Survey Questions}
\label{secttion:completesurvey} 

Questions related to taste for privacy:

\begin{enumerate}
    \item ``Are your social media profiles typically public or private?'' with response options ``Public, Mixed, Private.'' 
    \item ``Which platforms do you use?'' with response options ``Twitter, Instagram, Facebook, TikTok, Other.'' 
    \item ``In a typical month, how often do you use social media?'' with response options ``More than once a day, Once a day, A few times a week, Once a week, A few times a month, Once a Month, I use social media but less than once a month, I do not use social media.'' 
    \item ``How comfortable are you with sharing your personally identifiable data with a'' and we list 17 institutions as response options ``Government official, Police Officer, Friend, Relative, Employer, Medical Professional, Financial Institution, Neighbor, Acquaintance, Co-worker, School, Researcher, Social media platform, Non-profit, Company where you are a customer, Company where you are not a customer, stranger.'' Participants rated each institution on a 7-point Likert scale, where 1 = Very comfortable, 2 = Comfortable, 3 = Slightly Comfortable, 4 = Neutral, 5 = Slightly Uncomfortable, 6 = Uncomfortable, 7 = Very uncomfortable. 
\end{enumerate}

Questions related to Adverse Childhood Experiences (ACEs): 

\begin{enumerate}
    \item ``Did you live with anyone who was depressed, mentally ill, or suicidal?'' with response options ``No, Yes, Don't know.''
    \item ``Did you live with anyone who was a problem drinker or alcoholic?'' with response options ``No, Yes, Don't know.''
    \item ``Did you live with anyone who used illegal street drugs or who abused prescription medications?'' with response options ``No, Yes, Don't know.''
    \item ``Did you live with anyone who served time or was sentenced to serve time in a prison, jail, or other correctional facility?'' with response options ``No, Yes, Don't know.''
    \item ``Were your parents separated or divorced?'' with response options ``No, Yes, Parents not married, Don't know.''
    \item ``How often did your parents or adults in your home ever slap, hit, kick, punch or beat each other up?'' with response options ``Never, Once, More than once, Don't know.''
    \item ``Not including spanking, (before age 18), how often did a parent or adult in your home ever hit, beat, kick, or physically hurt you in any way?'' with response options ``Never, Once, More than once, Don't know.''
    \item ``How often did a parent or adult in your home ever swear at you, insult you, or put you down?'' with response options ``Never, Once, More than once, Don't know.''
    \item ``How often did anyone at least 5 years older than you or an adult, ever touch you sexually?'' with response options ``Never, Once, More than once, Don't know.''
    \item ``How often did anyone at least 5 years older than you or an adult, try to make you touch them sexually?'' with response options ``Never, Once, More than once, Don't know.''
    \item ``How often did anyone at least 5 years older than you or an adult, force you to have sex?'' with response options ``Never, Once, More than once, Don't know.''
    \item ``For how much of your childhood was there an adult in your household who made you feel safe and protected?'' with response options ``never, a little of the time, some of the time, most of the time, or all of the time.''
    \item ``For how much of your childhood was there an adult in your household who tried hard to make sure your basic needs were met?'' with response options ``never, a little of the time, some of the time, most of the time, or all of the time.''
\end{enumerate}

\section{Supplementary Figures and Tables}
\label{secttion:demographics} 

\begin{table*}[ht]
\centering
\caption{Demographics for the 2 baseline surveys.}
\label{table:demographicstable}
\begin{tabular}{lrr}
\toprule
Variable & No. responses over 2 baseline surveys & Percent \\
\midrule
Gender Male & 231 & 23.52 \\
Gender Female & 604 & 61.51 \\
Gender Non-binary & 147 & 14.97 \\
Race White & 741 & 78.66 \\
Race Multi-racial & 107 & 37.94 \\
Race Person of Color & 94 & 9.09 \\
First Generation Yes & 81 & 8.62 \\
First Generation  No & 859 & 91.38 \\
Sexual Orientation Straight & 441 & 50.88 \\
Sexual Orientation Bi-sexual & 206 & 23.76 \\
Sexual Orientation Homosexual & 78 & 9.00 \\
Sexual Orientation Other & 142 & 16.38 \\
Therapy Yes & 641 & 68.12 \\
Therapy No & 300 & 31.88 \\
\bottomrule
\end{tabular}
\end{table*}

\begin{table*}[ht]
\centering
\caption{Adverse Childhood Experiences collected over 4 surveys.}
\label{table:acestables}
\begin{tabular}{lrr}
\toprule
ACE & No. responses over 4 surveys & Percent \\
\midrule
Lived with someone with mental illness & 723 & 45.27 \\ 
Lived with someone with alcoholism & 361 & 22.60\\
Lived with someone who used drugs & 140 & 8.77 \\
Lived with someone who was incarcerated & 75 & 4.70 \\
Parents divorced or separated & 332 & 20.79 \\
Parental domestic violence toward each other & 205 & 12.84 \\
Parental physical abuse & 264 & 16.53 \\ 
Parental verbal abuse & 764 & 47.84 \\ 
Sexual touching by older person/adult & 103 & 6.45 \\
Coerced sexual contact by older person/adult & 77 & 4.82 \\
Forced sex by older person/adult & 25 & 1.57 \\
Adult who made you feel safe/protected (Never/a little of the time) & 82 & 5.13 \\ 
Adult who met your basic needs (Never/a little of the time) & 48 & 3.01 \\
Student who reported no ACE & 353 & 22.10 \\
Students who reported at least one ACE & 1244 & 77.90 \\
\bottomrule
\end{tabular}
\end{table*}

\label{sec:likertinstdescriptivestats}

\begin{table*}[ht]
\caption{Descriptive statistics for comfort/privacy variables by institution type}
\begin{tabular}{lrrrrrrr}
\hline
Institution & Mean & Std & Min & 25\% & 50\% & 75\% & Max \\
\hline
Friend & 2.232246 & 1.168967 & 1.0 & 1.0 & 2.0 & 3.0 & 7.0 \\

Medical & 2.600580 & 1.351796 & 1.0 & 2.0 & 2.0 & 3.0 & 7.0 \\

Relative & 2.657962 & 1.330732 & 1.0 & 2.0 & 2.0 & 3.0 & 7.0 \\

Researcher & 3.294416 & 1.321804 & 1.0 & 2.0 & 3.0 & 4.0 & 7.0 \\

Employer & 3.602397 & 1.424148 & 1.0 & 3.0 & 3.0 & 5.0 & 7.0 \\

School & 3.757158 & 1.402538 & 1.0 & 3.0 & 4.0 & 5.0 & 7.0 \\

Financial & 3.817193 & 1.460395 & 1.0 & 3.0 & 4.0 & 5.0 & 7.0 \\

Non-Profit & 4.102611 & 1.348566 & 1.0 & 3.0 & 4.0 & 5.0 & 7.0 \\

Co-worker & 4.159898 & 1.505459 & 1.0 & 3.0 & 4.0 & 5.0 & 7.0 \\

Government & 4.249366 & 1.639615 & 1.0 & 3.0 & 4.0 & 5.0 & 7.0 \\

Neighbor & 4.335507 & 1.470665 & 1.0 & 3.0 & 4.0 & 5.0 & 7.0 \\

Police & 4.379310 & 1.714628 & 1.0 & 3.0 & 5.0 & 6.0 & 7.0 \\

Acquaintance & 4.403191 & 1.576447 & 1.0 & 3.0 & 4.0 & 6.0 & 7.0 \\

Company (customer) & 4.716618 & 1.433800 & 1.0 & 4.0 & 5.0 & 6.0 & 7.0 \\

Platform & 4.738752 &1.557827 & 1.0 & 4.0 & 5.0 & 6.0 & 7.0 \\

Company (not a customer) & 5.519768 & 1.394861 & 1.0 & 5.0 & 6.0 & 6.0 & 7.0 \\

Stranger & 5.866159 & 1.383556 & 1.0 & 5.0 & 6.0 & 7.0 & 7.0 \\
\hline
\end{tabular}
\Description{}
\label{tab:descriptive_stats}
\end{table*}

\label{secttion:Meandiffplots} 

\begin{figure*}
    \centering
    \includegraphics[width=.3\linewidth]{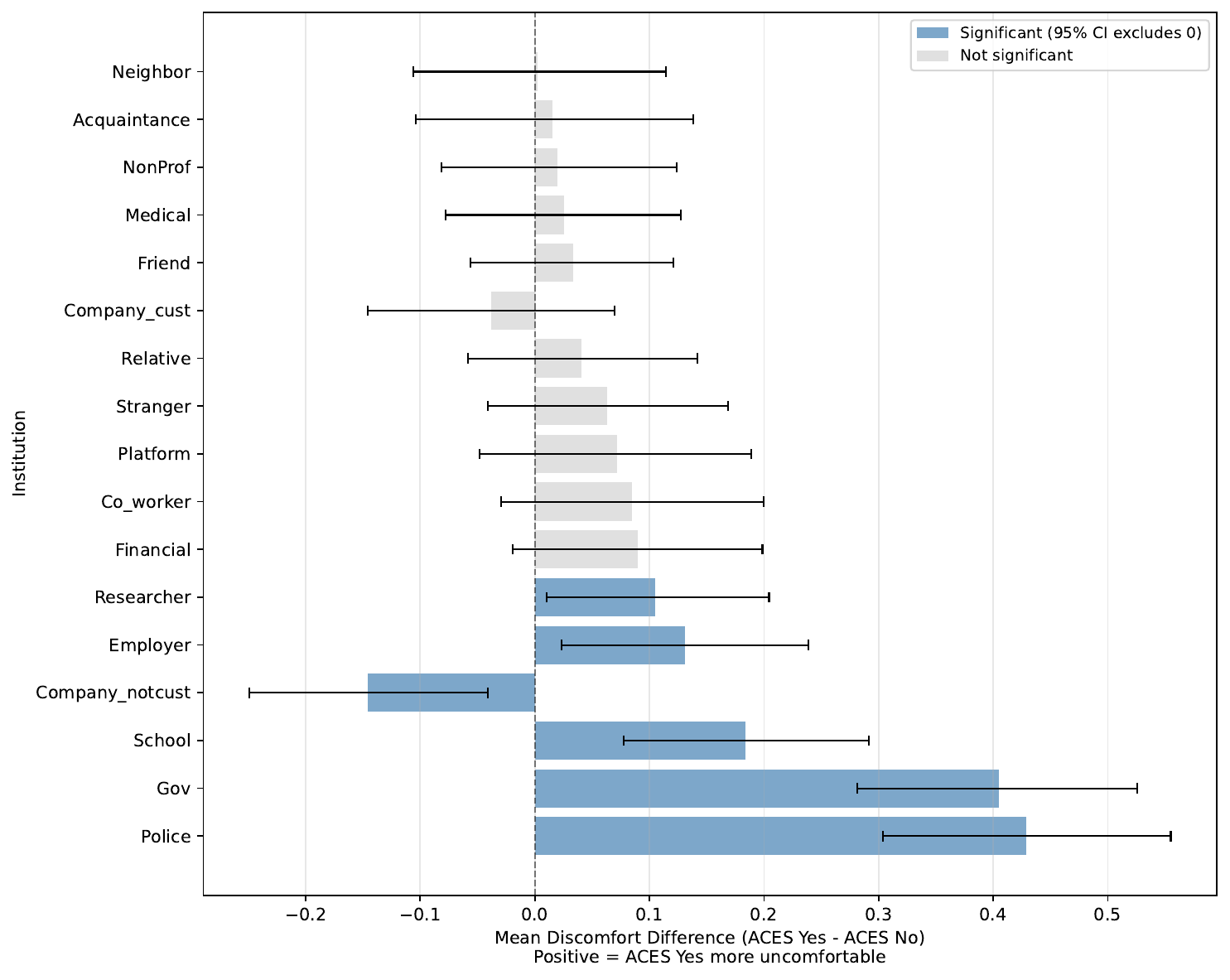}
    \includegraphics[width=.3\linewidth]{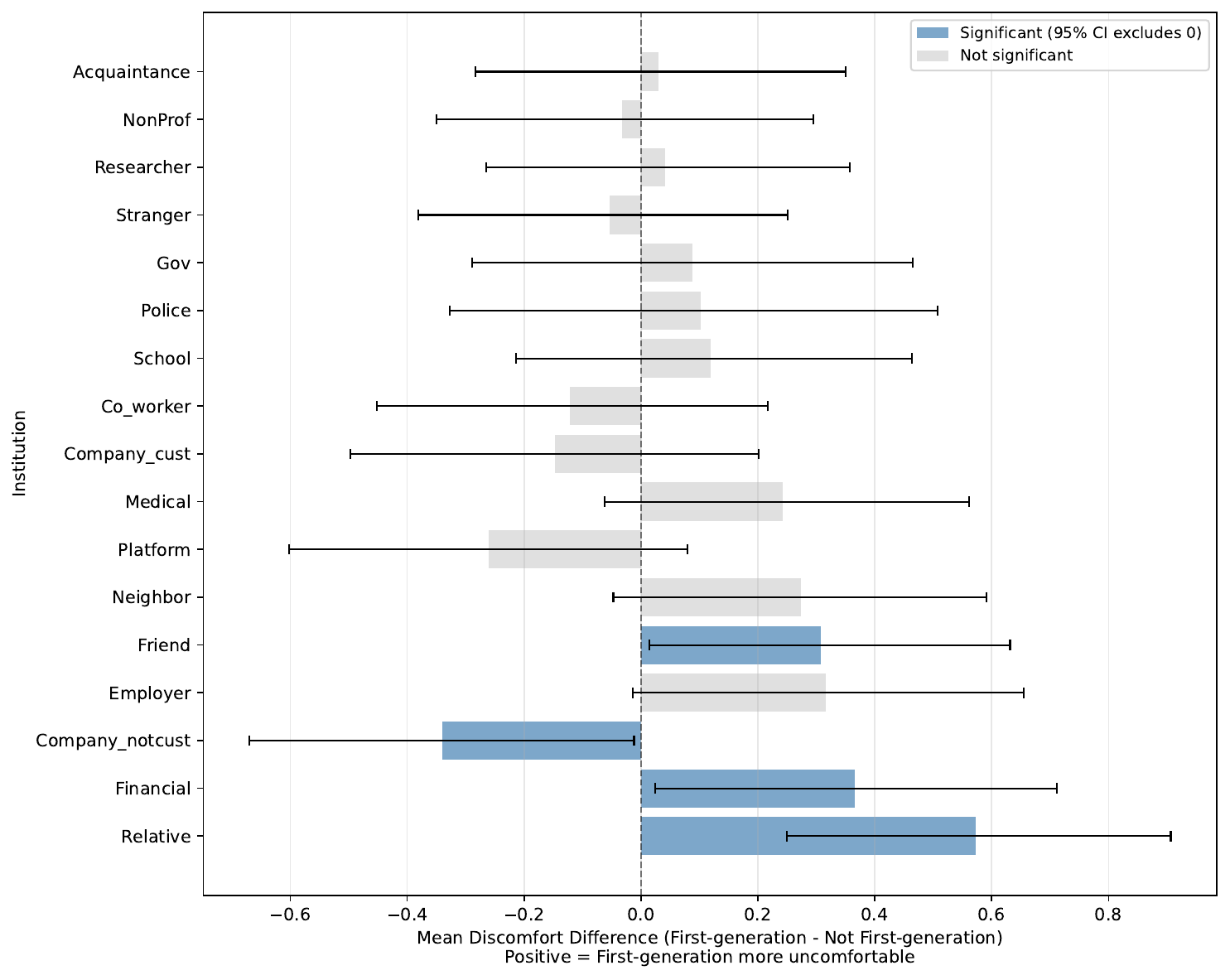}
    \includegraphics[width=.3\linewidth]{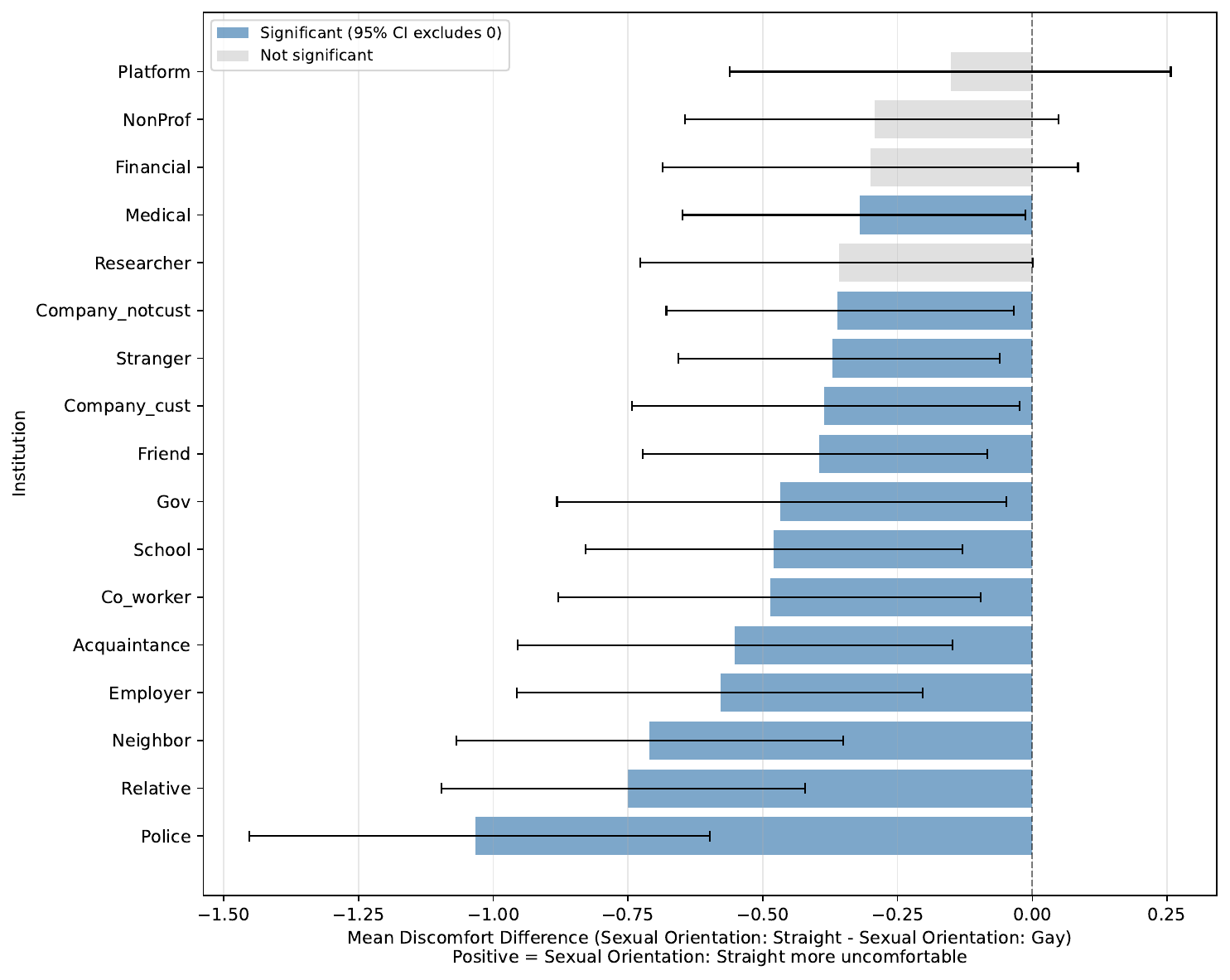}
    \includegraphics[width=.3\linewidth]{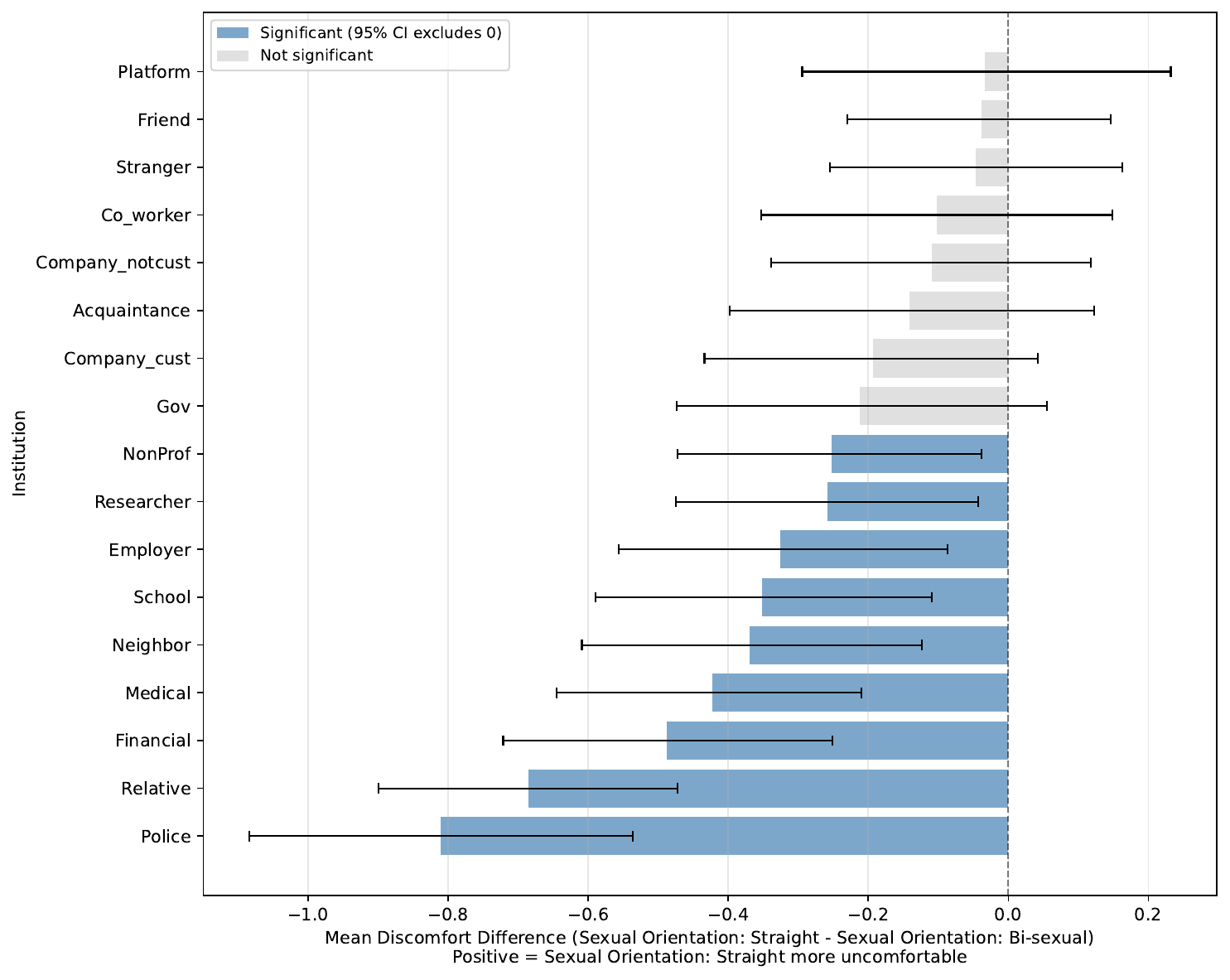}
    \includegraphics[width=.3\linewidth]{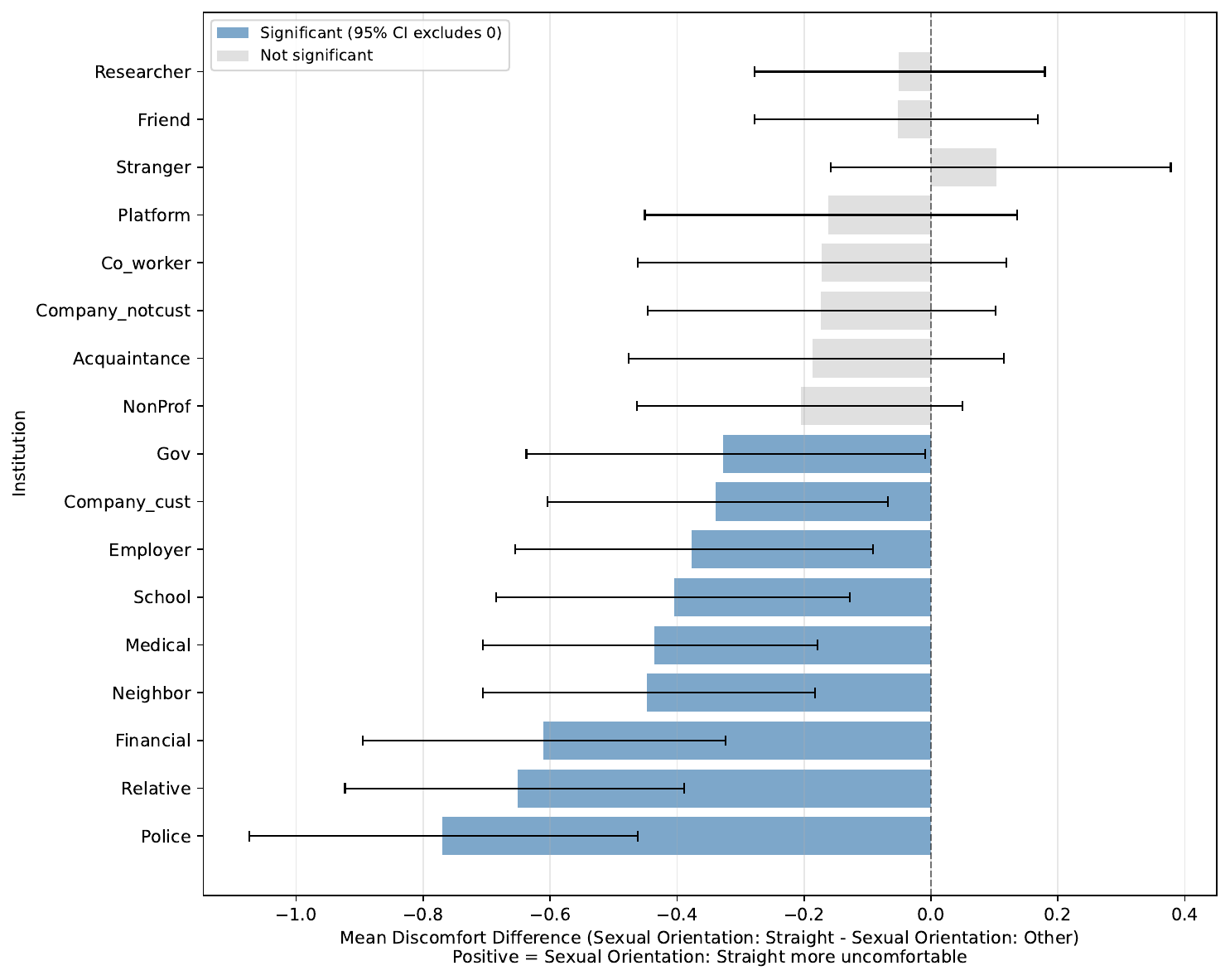}
    \includegraphics[width=.3\linewidth]{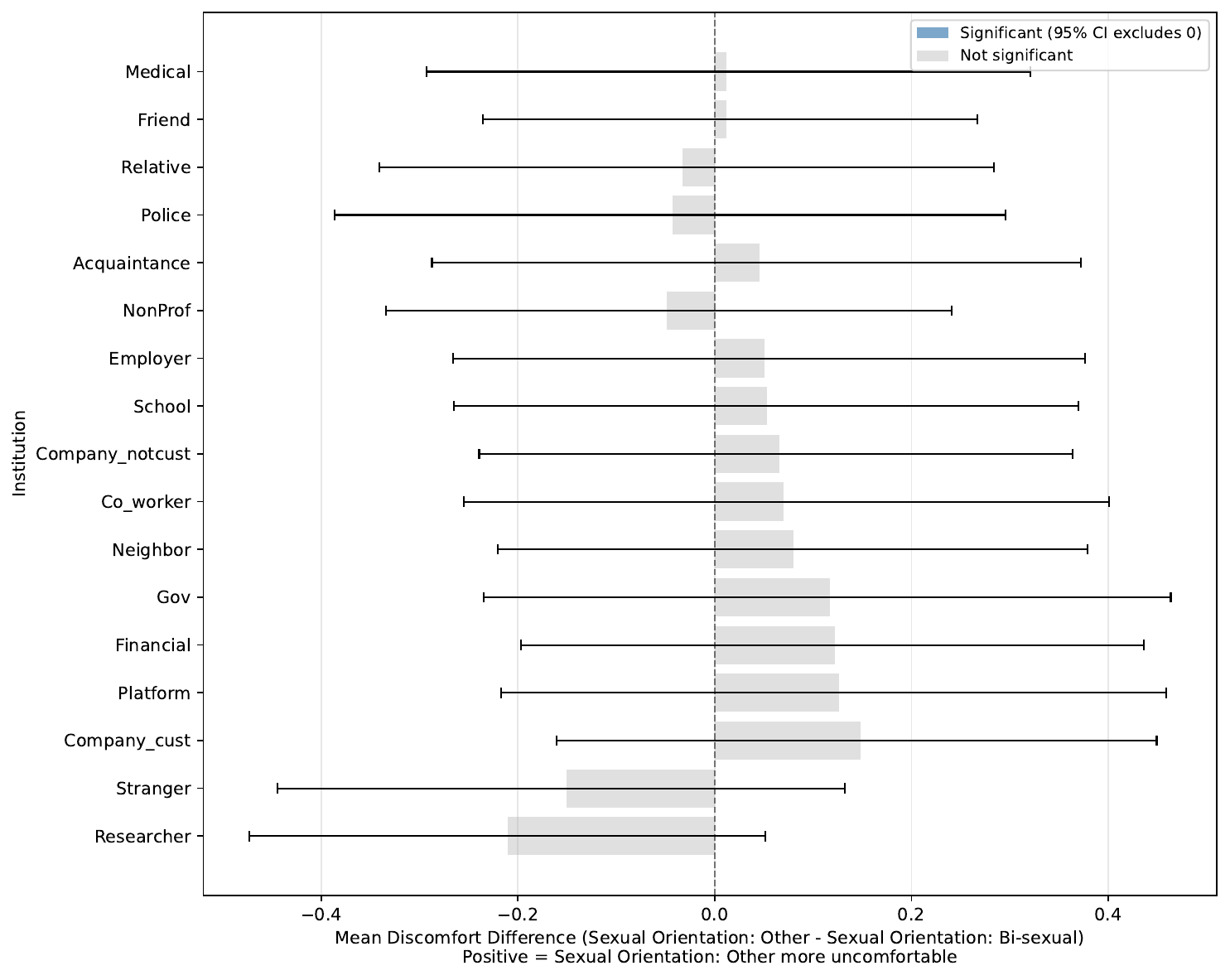}
    \includegraphics[width=.3\linewidth]{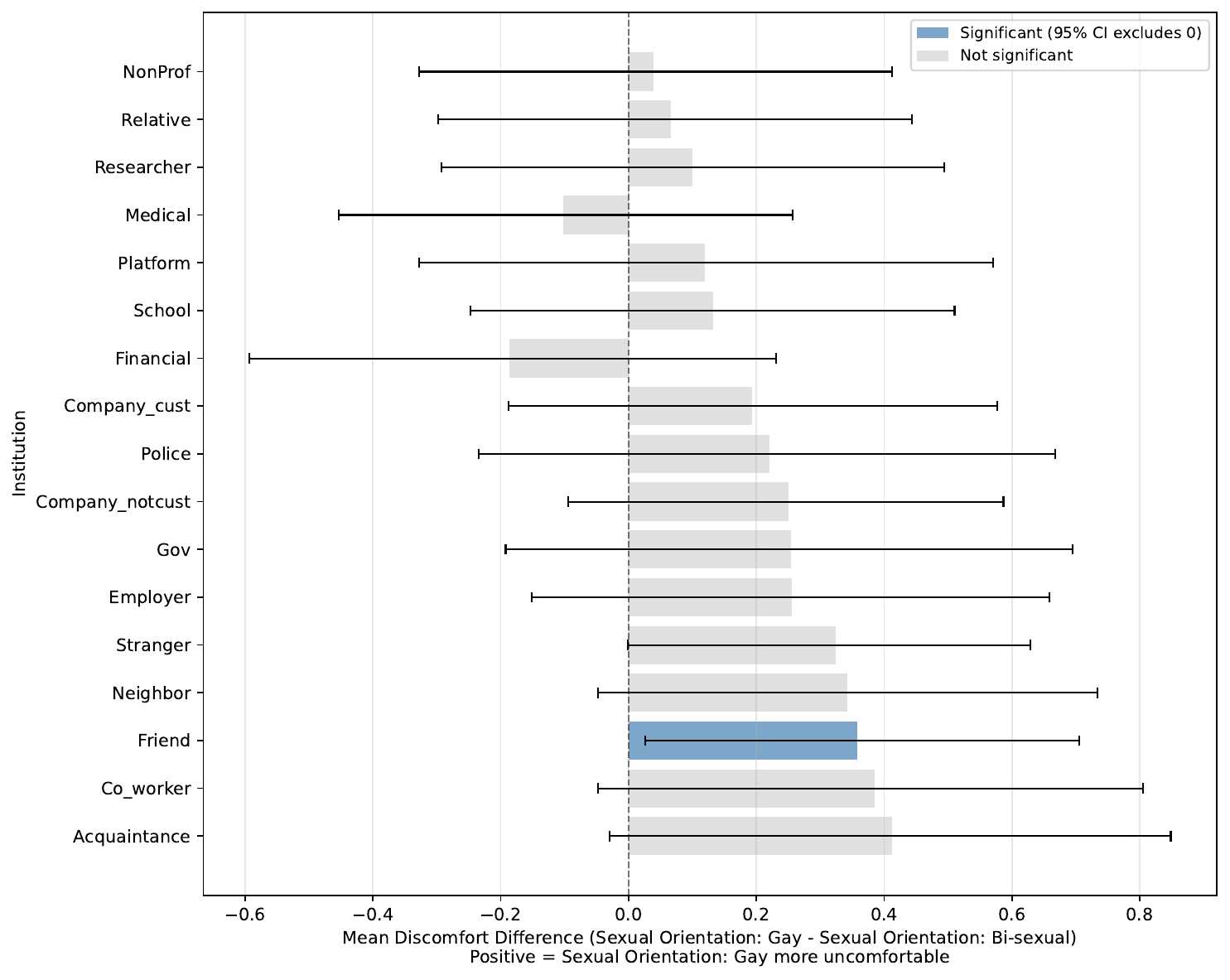}
    \includegraphics[width=.3\linewidth]{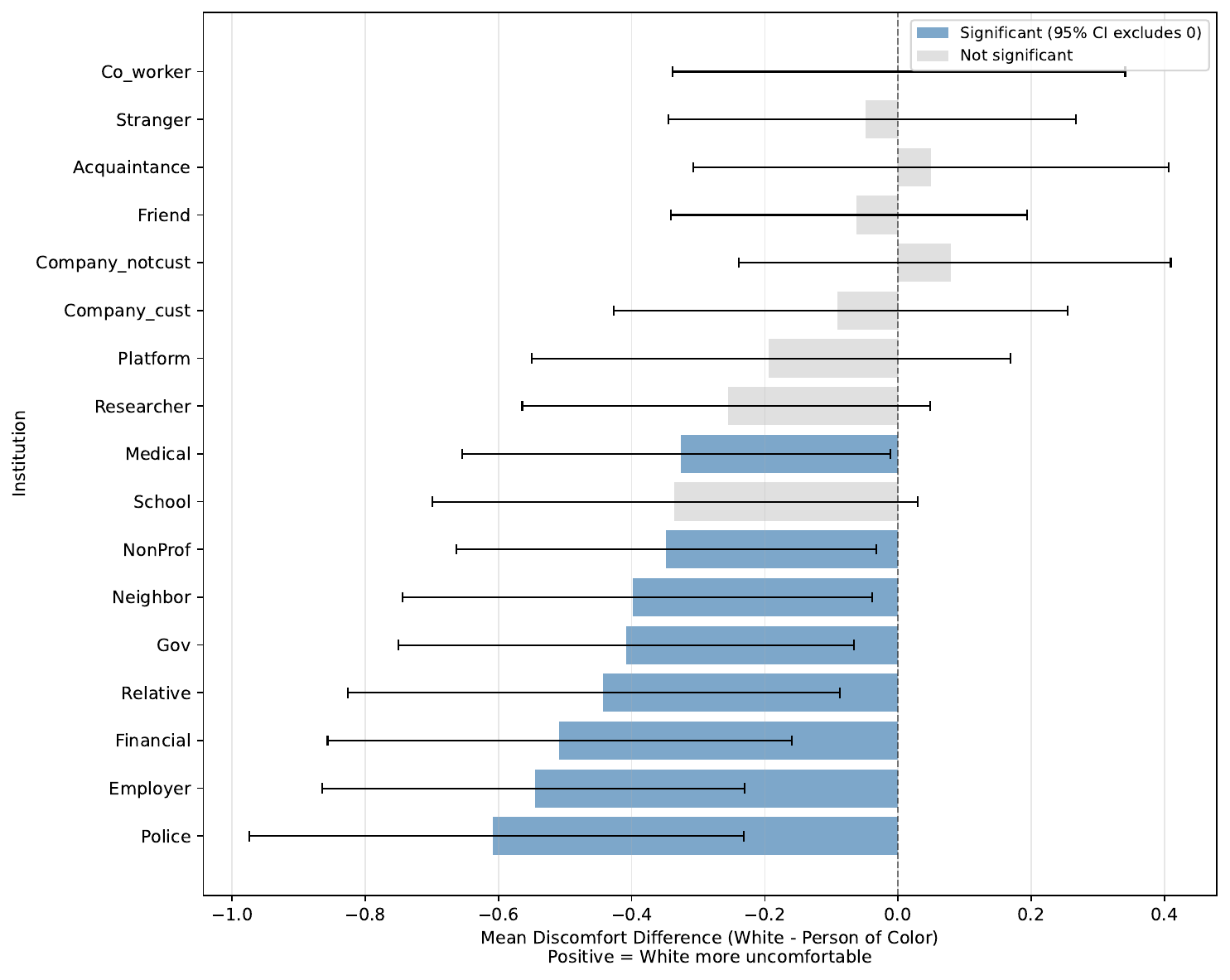}
    \includegraphics[width=.3\linewidth]{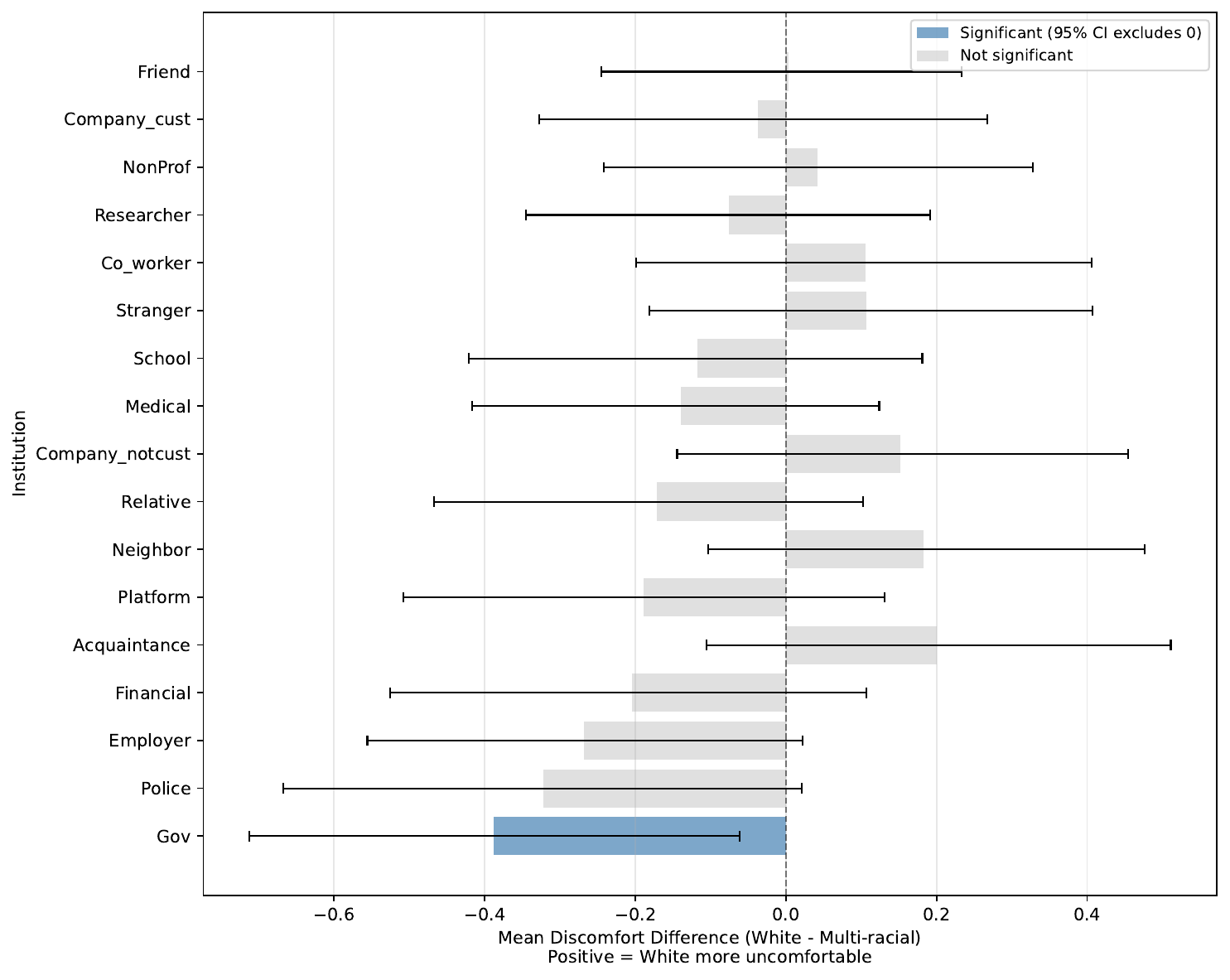}
    \includegraphics[width=.3\linewidth]{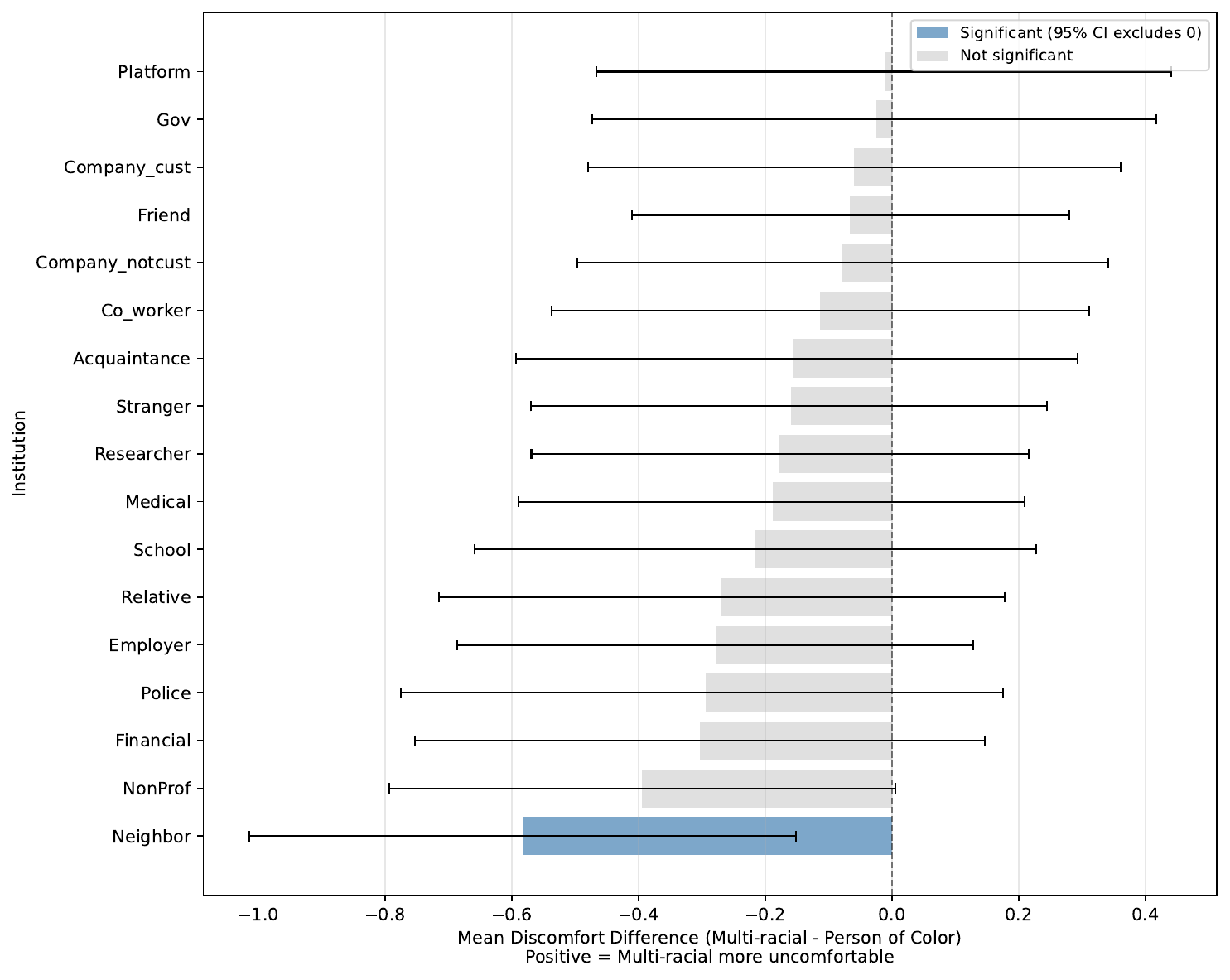}  
        \caption{Differences in mean discomfort sharing PII across demographic groups. Forest plots showing bootstrapped mean differences in discomfort (with 95\% confidence intervals) for 17 institutional contexts across three demographic comparisons. Blue bars indicate significant differences (95\% CI excludes zero); gray bars indicate non-significant differences.}
    \Description{Three side-by-side forest plots comparing discomfort levels across demographic groups. All significant differences are marked in blue with confidence intervals excluding zero; non-significant differences appear in gray.}
\label{fig:therapymeandiffbootstrapped}
\end{figure*}

\label{secttion:ranksiffplots} 

\begin{figure*}
    \centering
    \includegraphics[width=.3\linewidth]{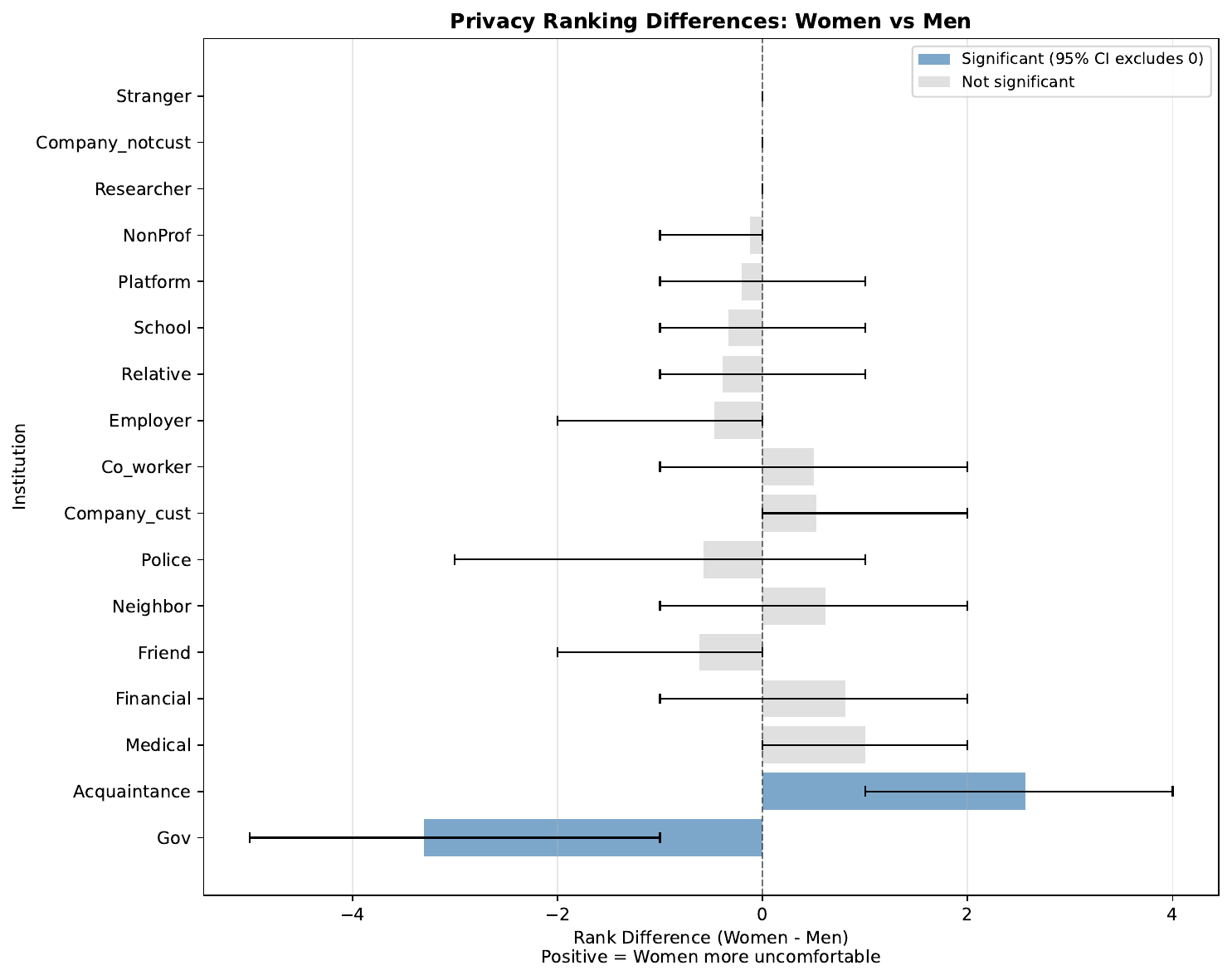}
    \includegraphics[width=.3\linewidth]{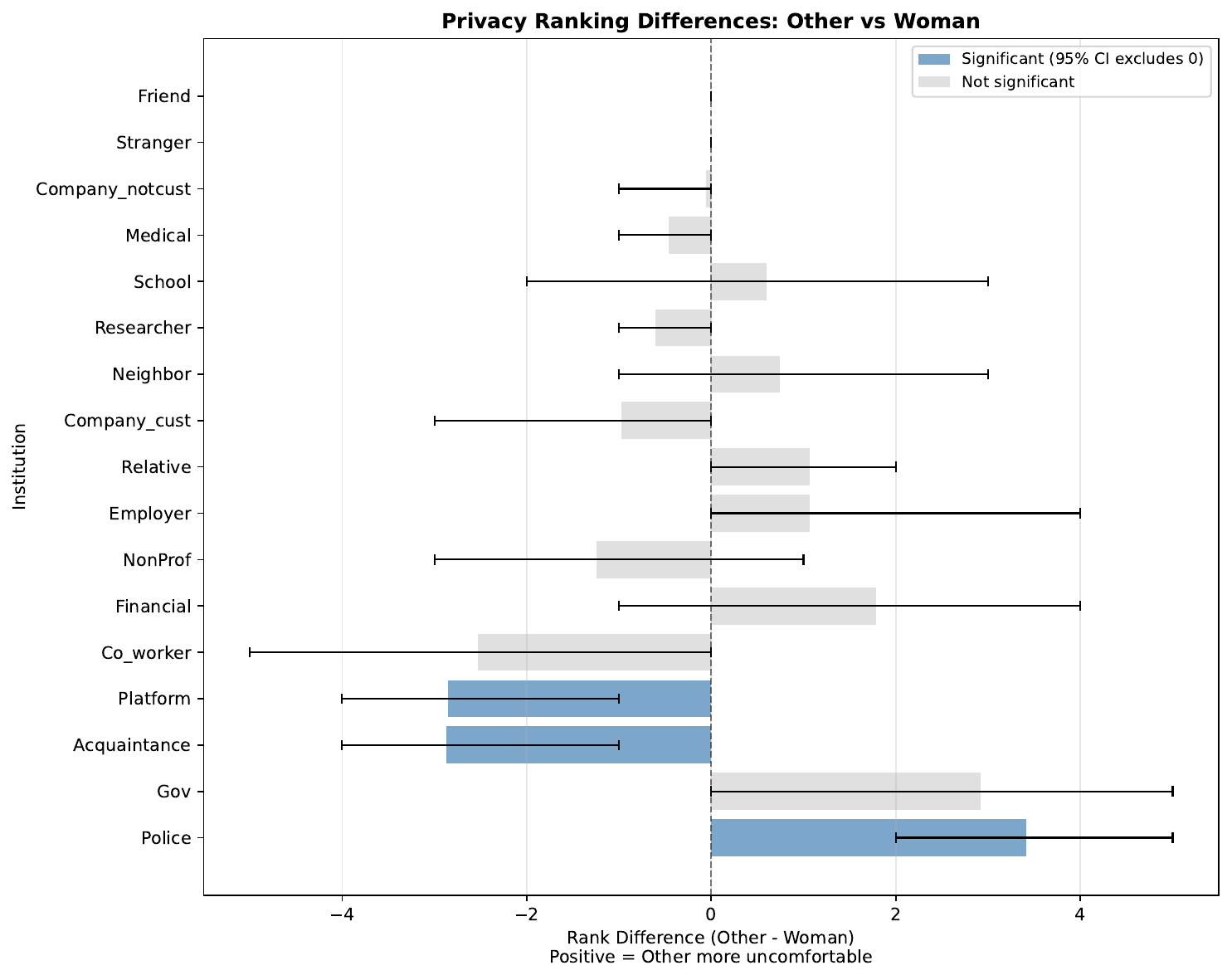}
    \includegraphics[width=.3\linewidth]{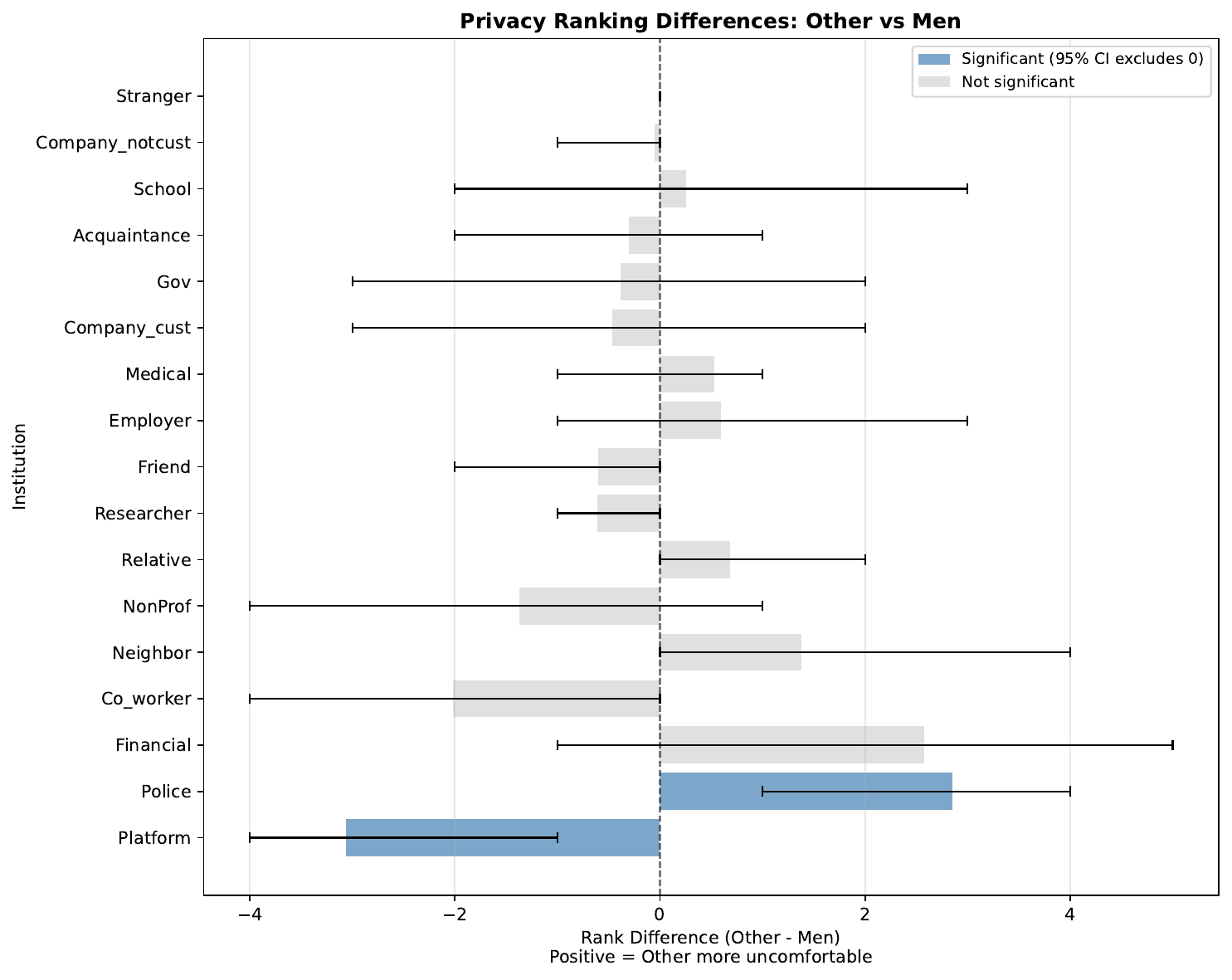}
    \includegraphics[width=.3\linewidth]{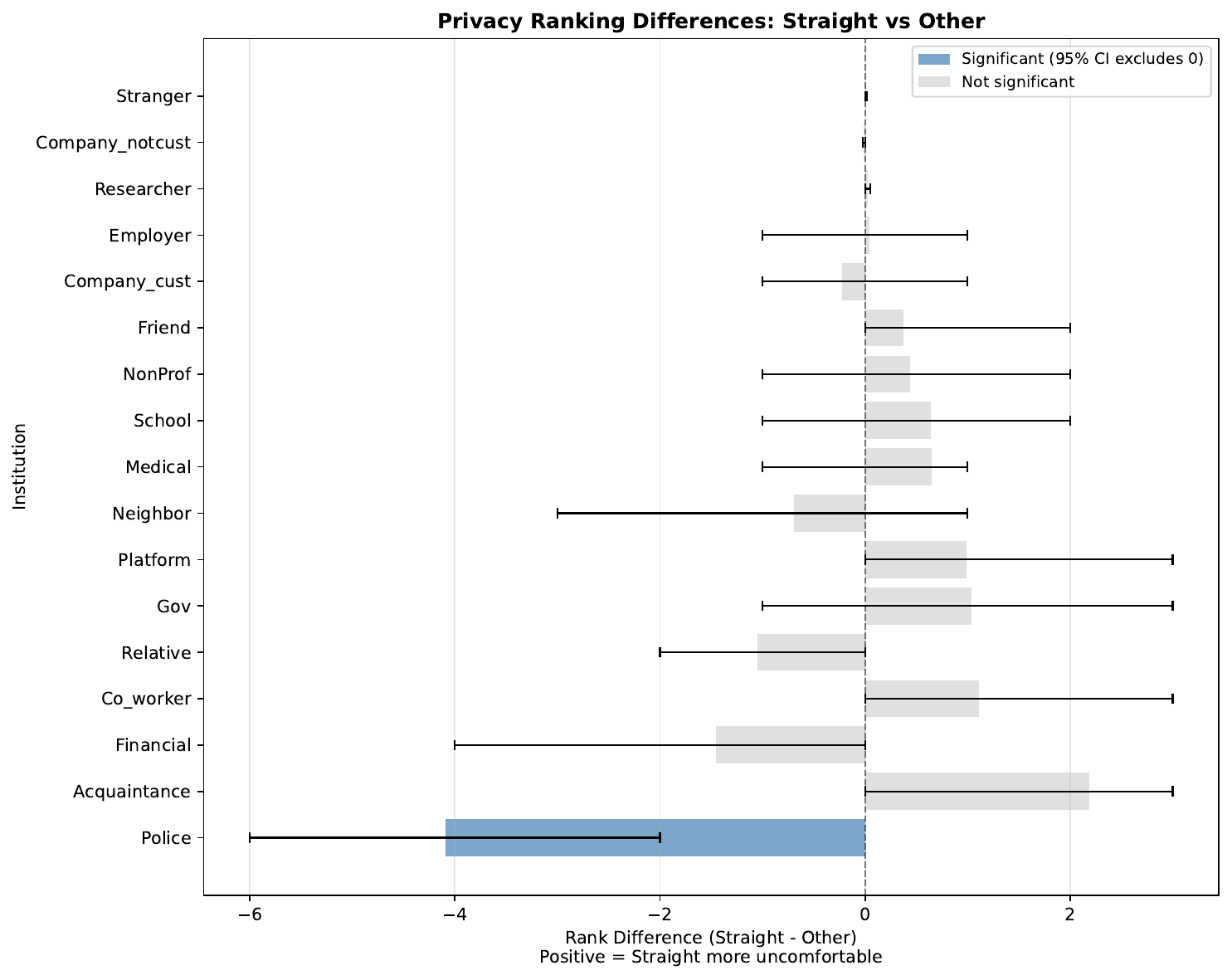}
    \includegraphics[width=.3\linewidth]{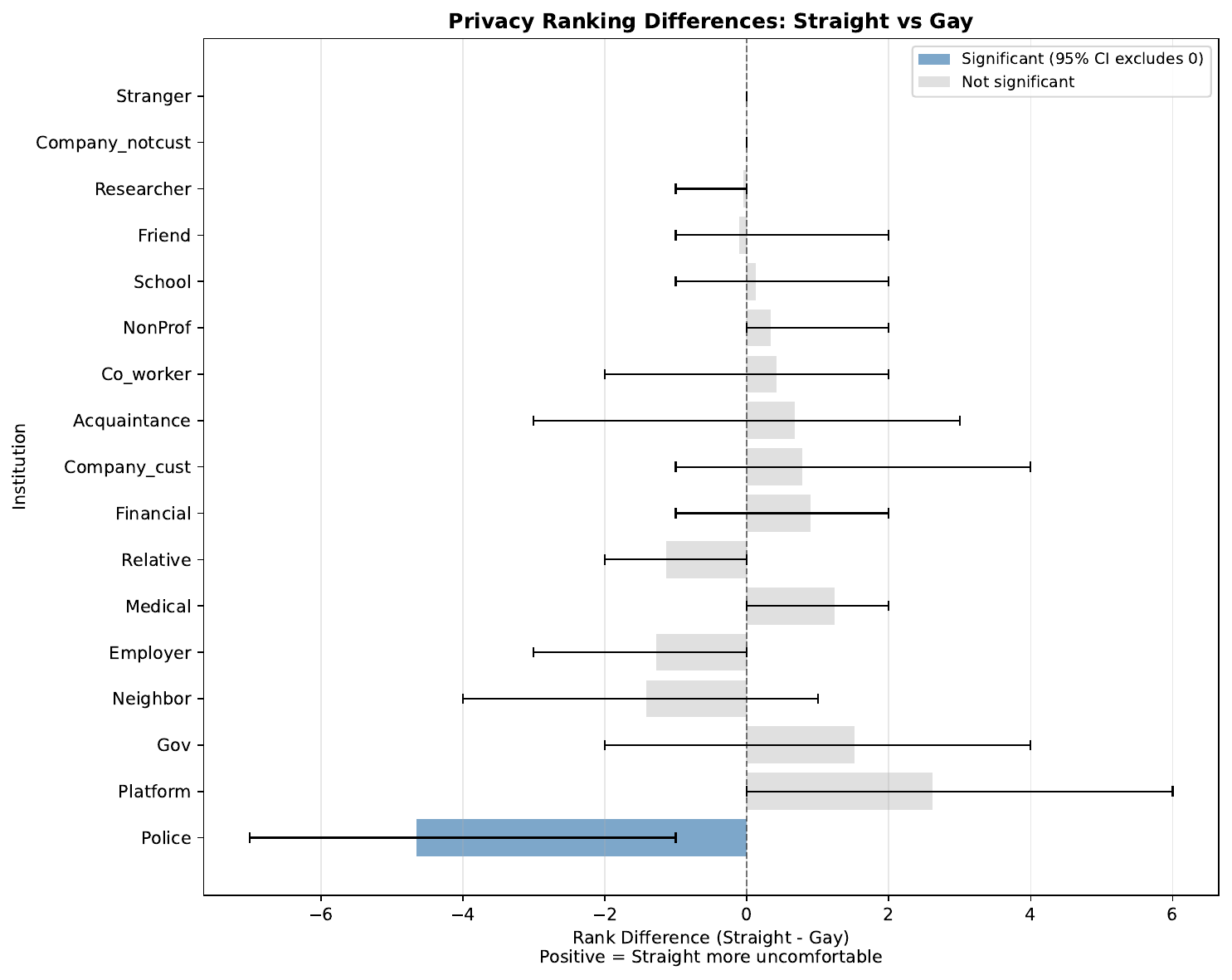}
    \includegraphics[width=.3\linewidth]{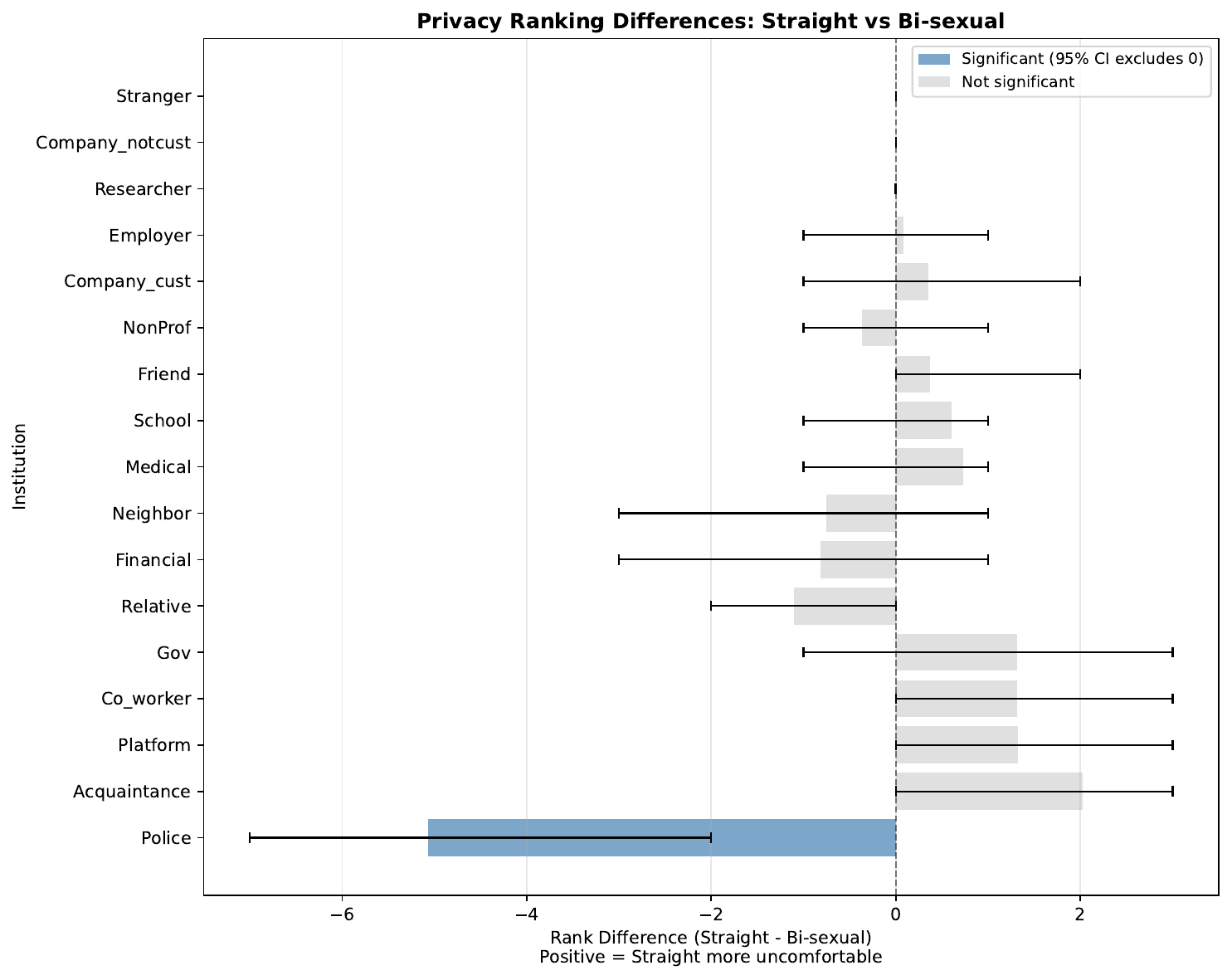}
    \includegraphics[width=.3\linewidth]{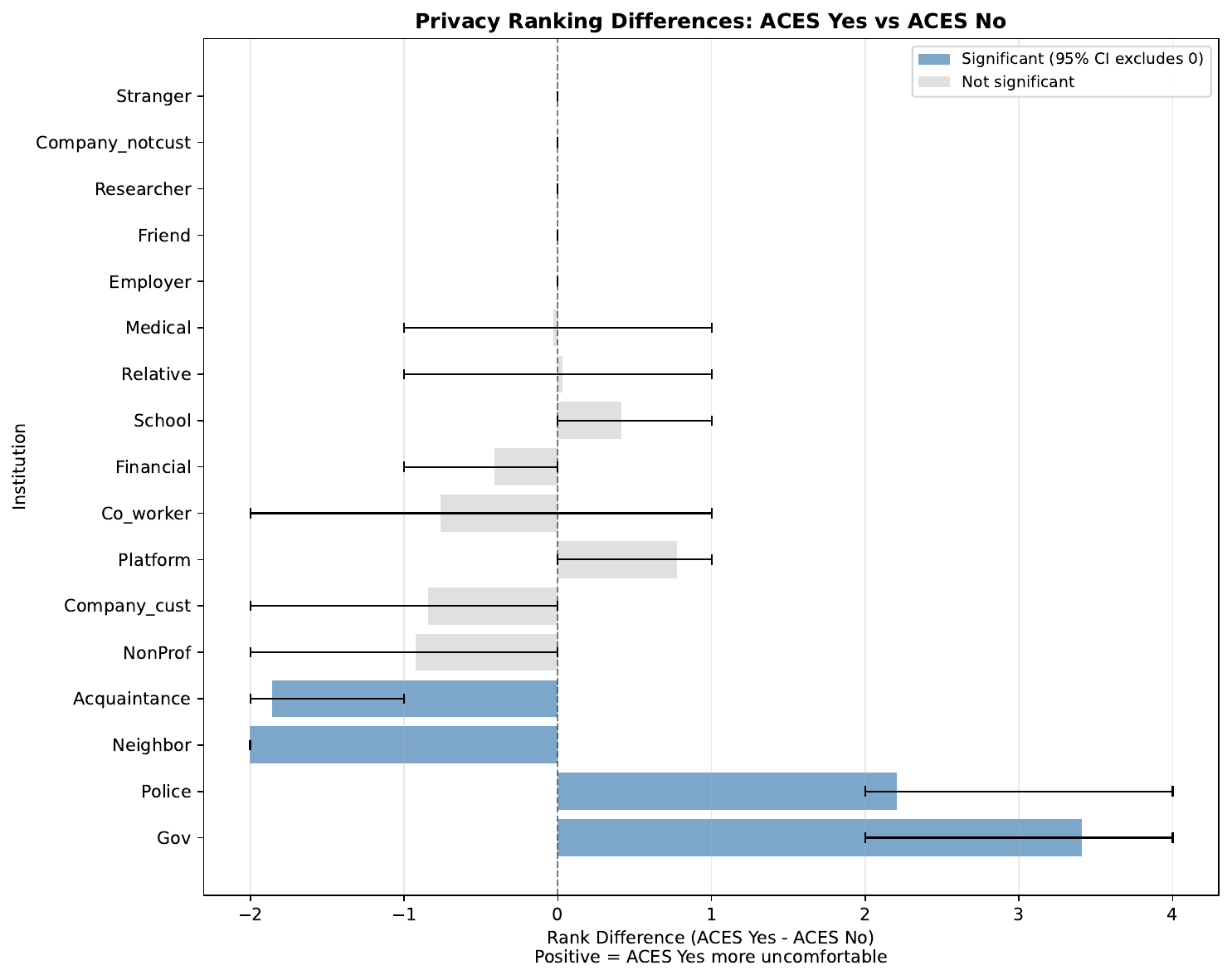}
    \caption{Privacy ranking differences across significant demographic comparisons. Each panel shows the difference in institution discomfort rankings between two demographic groups (Group 1 rank minus Group 2 rank), where positive values indicate institutions ranked as more uncomfortable by Group 1. Error bars represent 95\% confidence intervals from 10,000 bootstrap iterations. Blue bars indicate statistically significant differences (confidence intervals exclude zero); gray bars indicate non-significant differences.}
    \Description{Grid of 12 horizontal bar charts showing privacy ranking differences across demographic comparisons. Each panel displays approximately 18 institution types on the y-axis (including Police, Co-worker, Acquaintance, Gov, Neighbor, Relative, Financial, School, Friend, Medical, Company Customer, Non Profit, Researcher, Employer, Platform, Company Not A Customer, Stranger, and variations by context) and rank differences on the x-axis. A vertical dashed line at zero marks no difference. Each chart shows error bars representing 95\% confidence intervals, with blue bars indicating statistically significant differences and gray bars indicating non-significant differences. The pattern and number of significant differences varies across demographic comparisons, with some panels showing multiple blue bars and others showing primarily gray bars.}
    \label{fig:rankdiffbootstrapped}
\end{figure*}

\label{secttion:likertarcplots}

\begin{figure*}
    \centering
    \includegraphics[width=.3\linewidth]{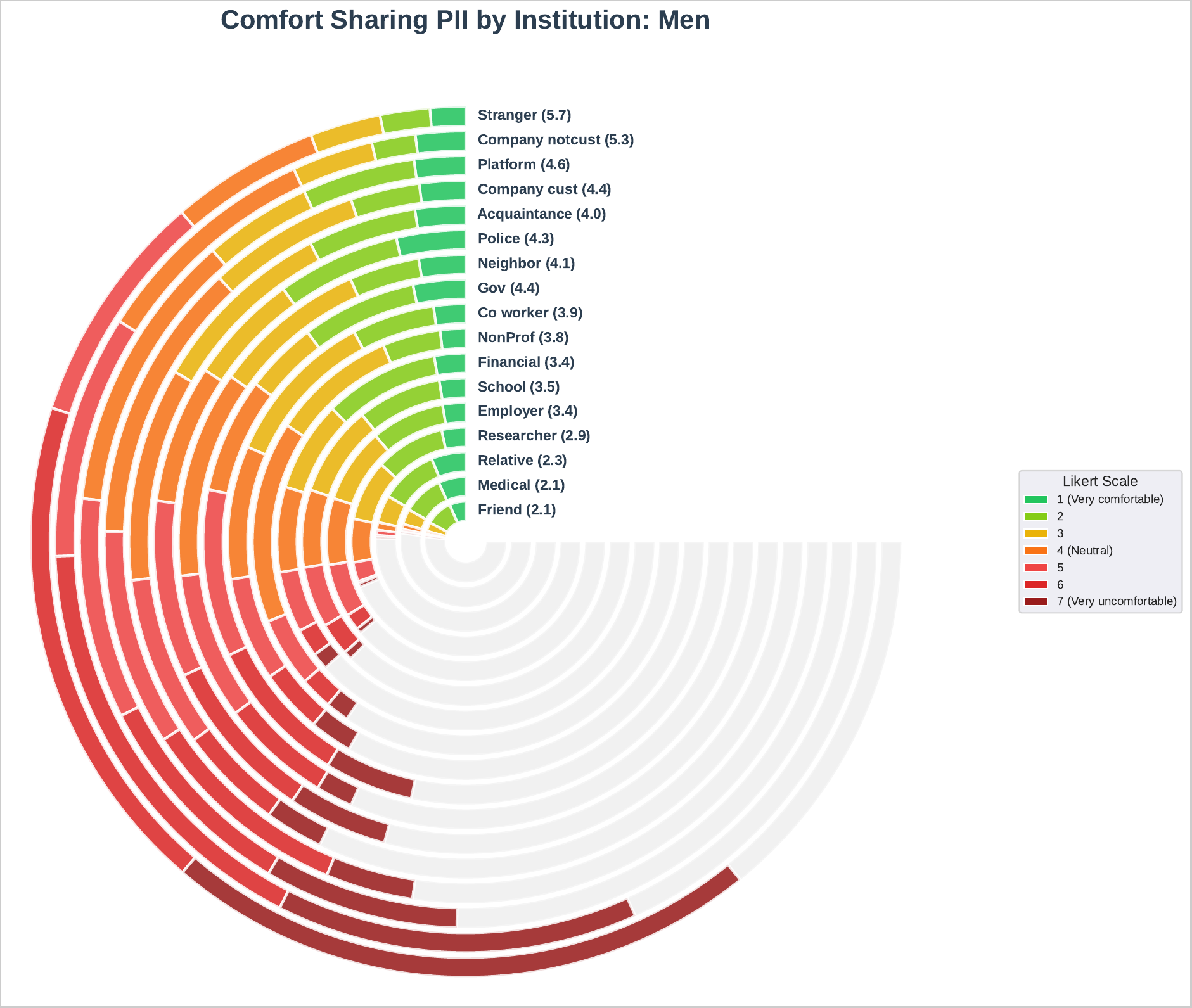}
    \includegraphics[width=.3\linewidth]{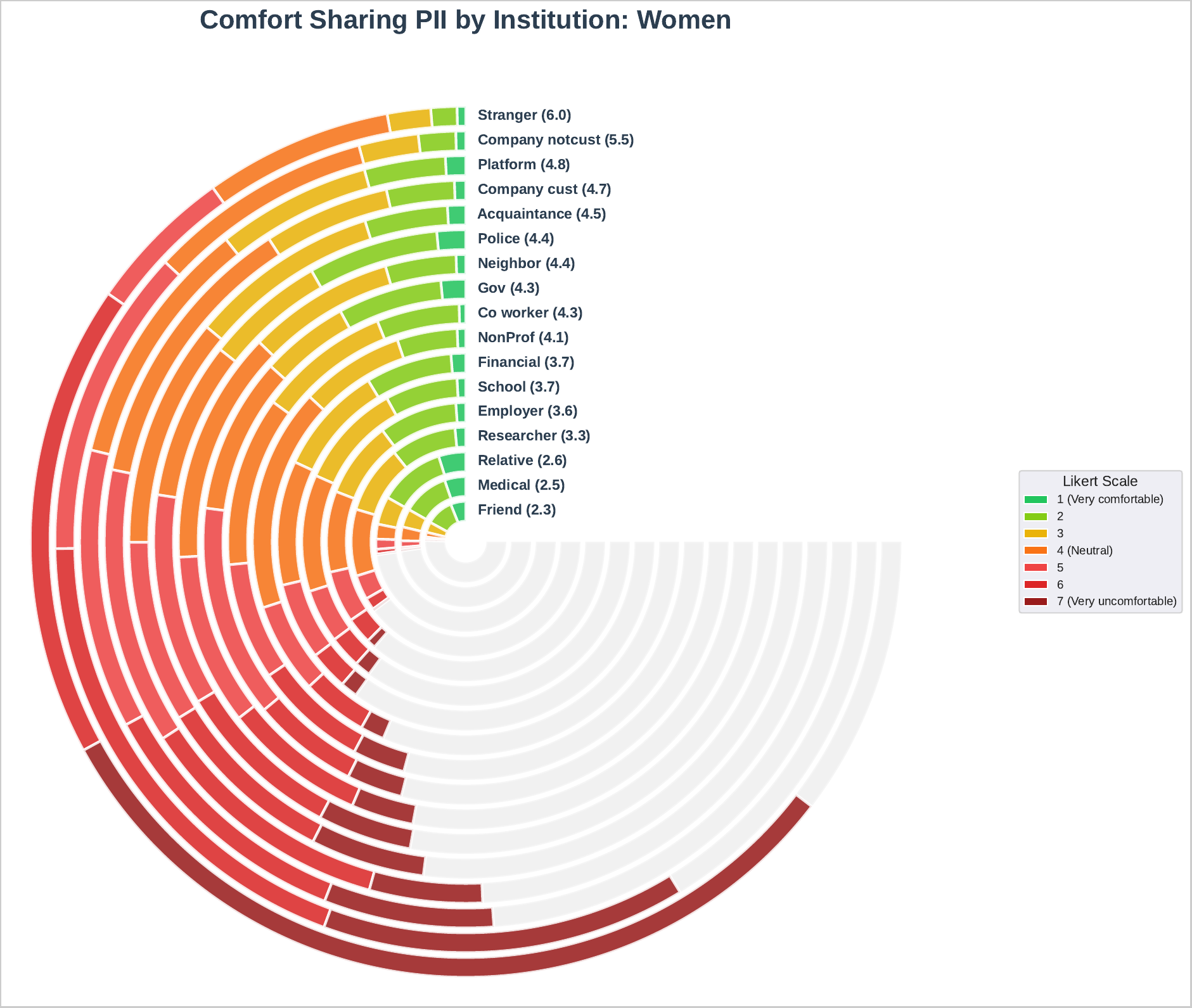}
    \includegraphics[width=.3\linewidth]{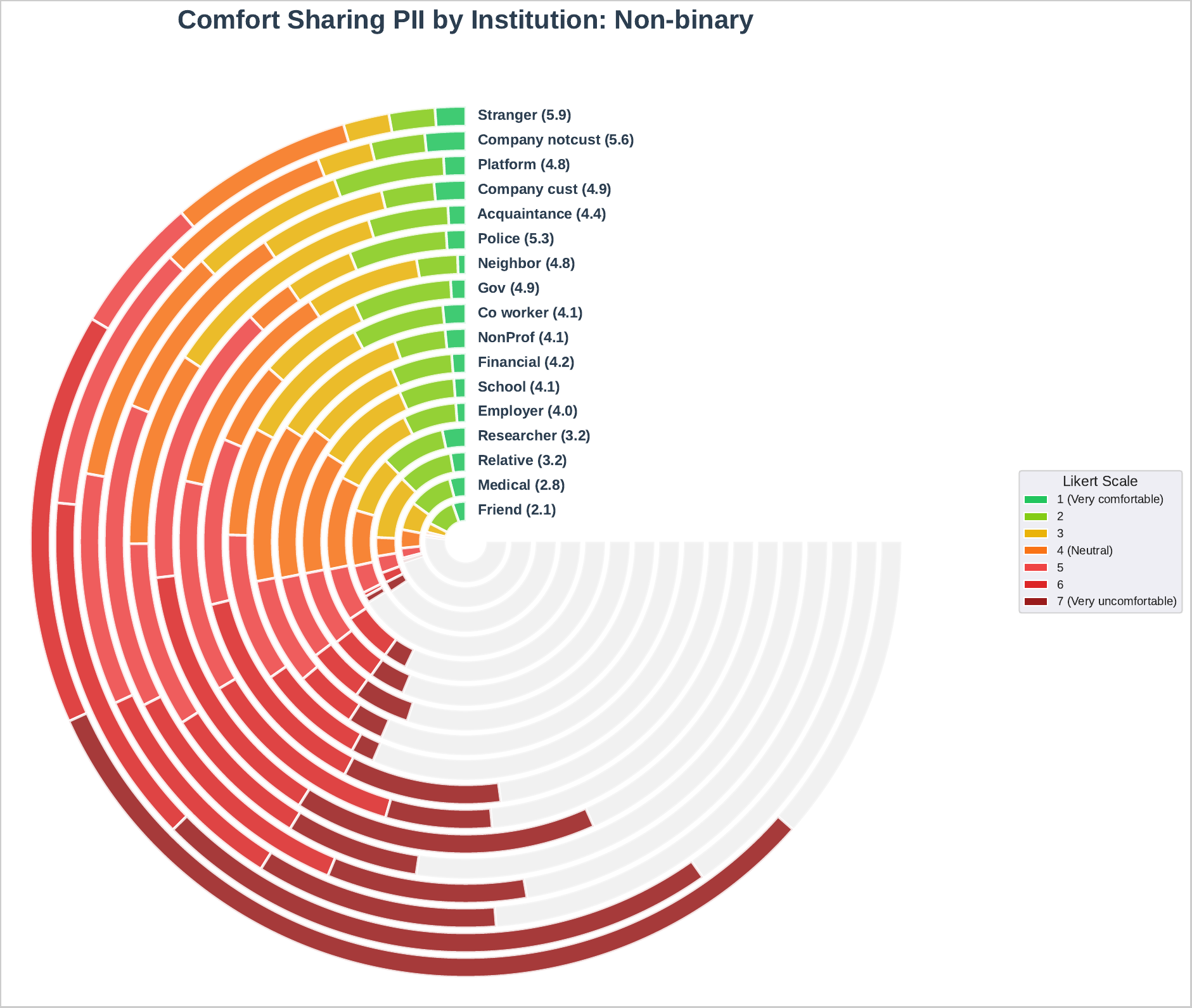}
    \caption{Comfort Sharing PII by Institution and Gender. Arc length represents mean discomfort level, ranging from 1 when very comfortable (green) to 7 when very uncomfortable (red). Colored segments show the full breakdown of Likert responses.}
    \Description{Circular stacked bar charts showing comfort levels for sharing personal information with 17 institutions by gender. Each institution is represented by a horizontal bar bent into an arc, divided into colored segments for the 7-point Likert scale. Green segments indicate comfort, yellow/orange neutral responses, and red segments discomfort.}
    \label{fig:arcdemplotsgender}
\end{figure*}

\begin{figure*}
    \centering
    \includegraphics[width=.3\linewidth]{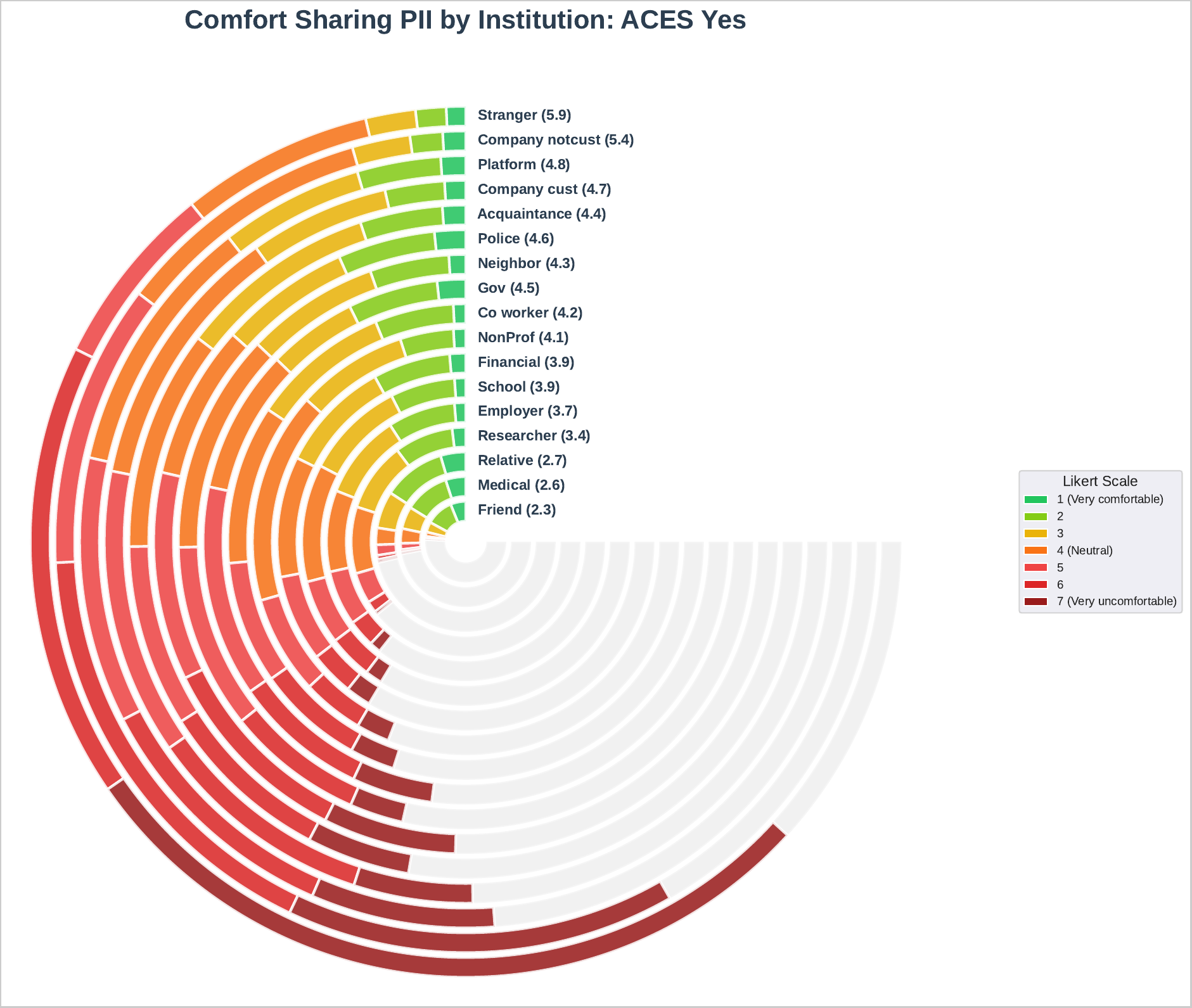}
    \includegraphics[width=.3\linewidth]{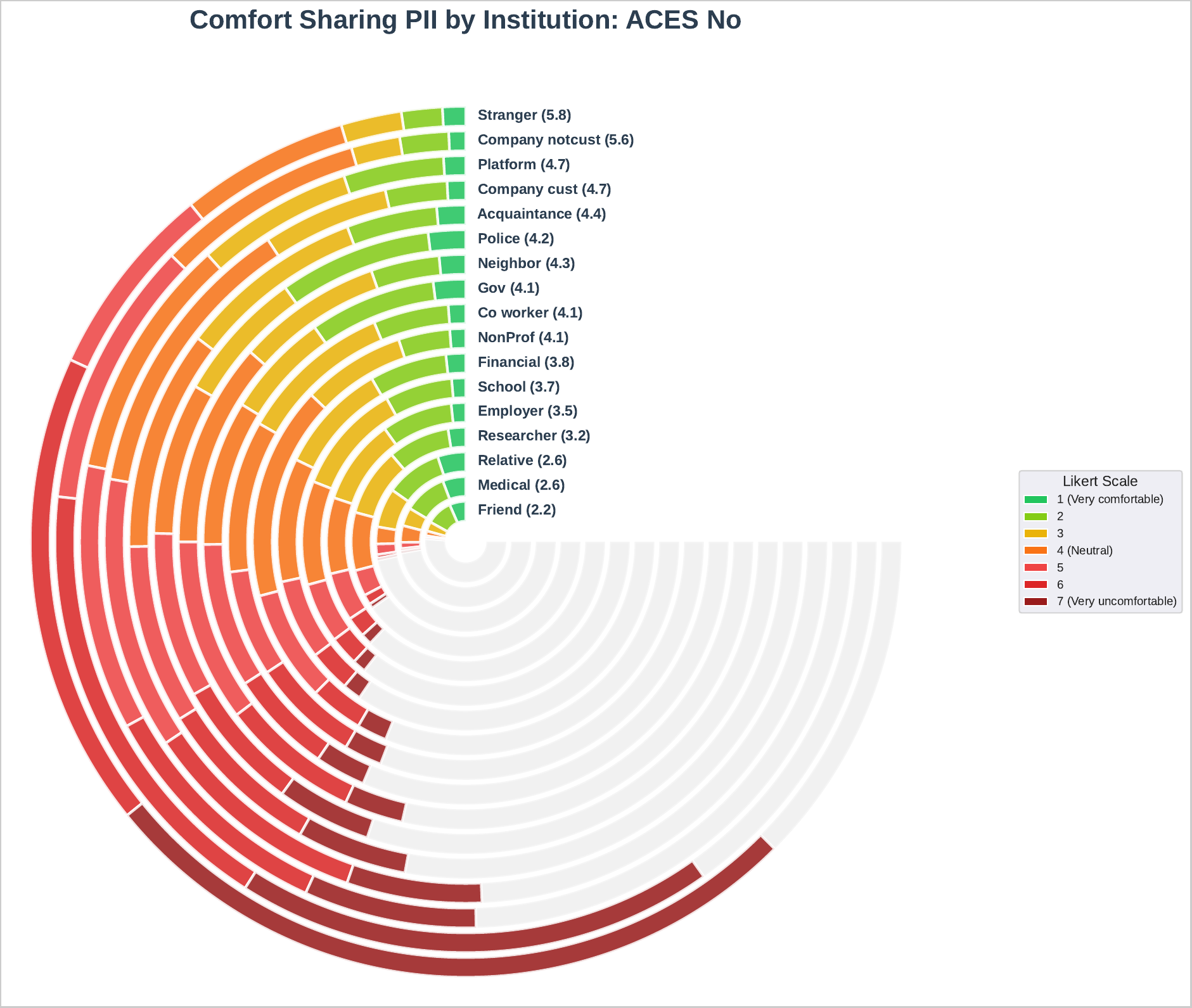}
    \caption{Comfort Sharing PII by Institution and Adverse Childhood Experiences. Arc length represents mean discomfort level, ranging from 1 when very comfortable (green) to 7 when very uncomfortable (red). Colored segments show the full breakdown of Likert responses.}
    \Description{Circular stacked bar chart showing comfort levels for sharing personal information with 17 institutions by ACEs. Each institution is represented by a horizontal bar bent into an arc, divided into colored segments for the 7-point Likert scale. Green segments indicate comfort, yellow/orange neutral responses, and red segments discomfort.}
    \label{fig:arcdemplotsACES}
\end{figure*}

\begin{figure*}
    \centering
    \includegraphics[width=.3\linewidth]{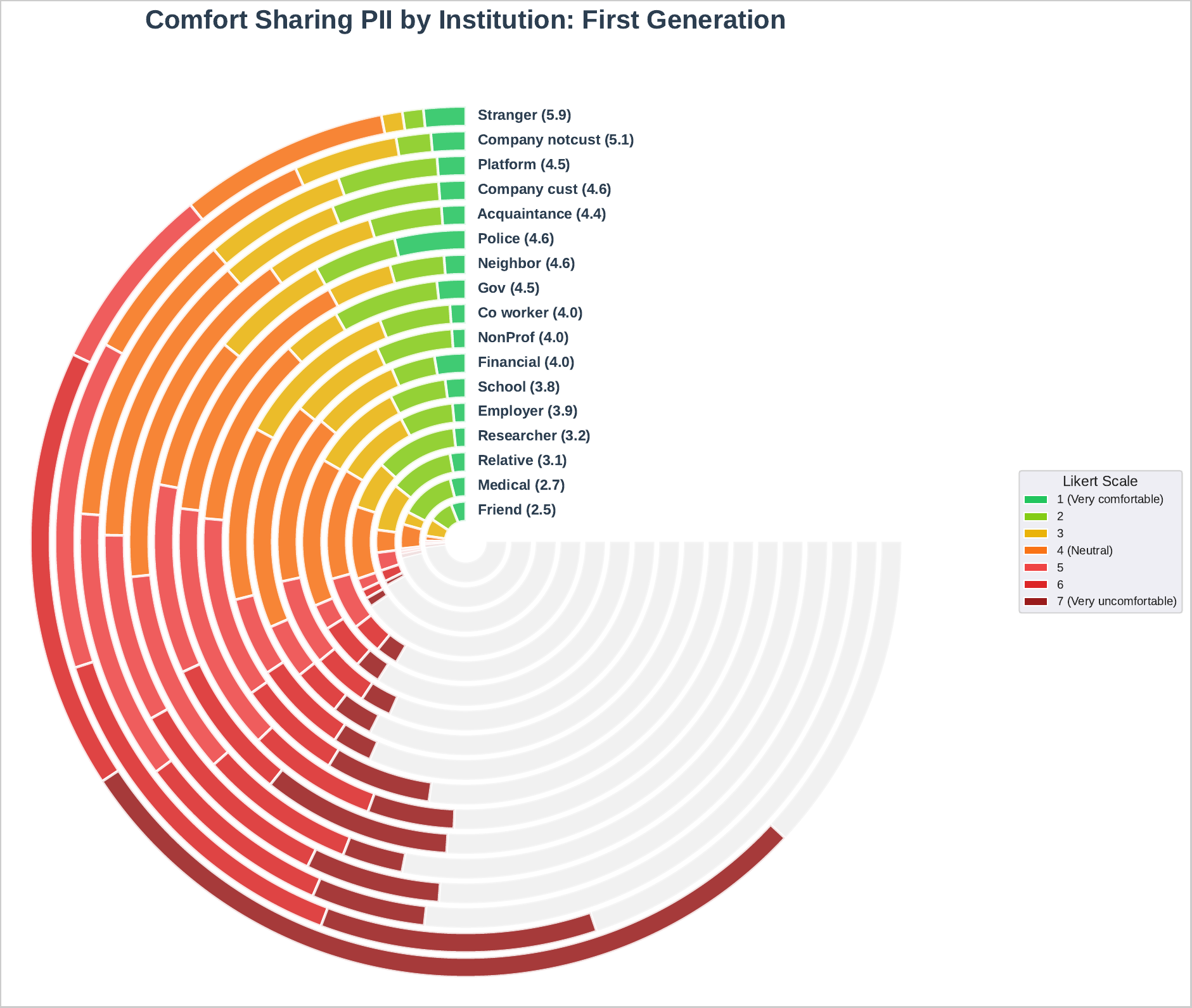}
    \includegraphics[width=.3\linewidth]{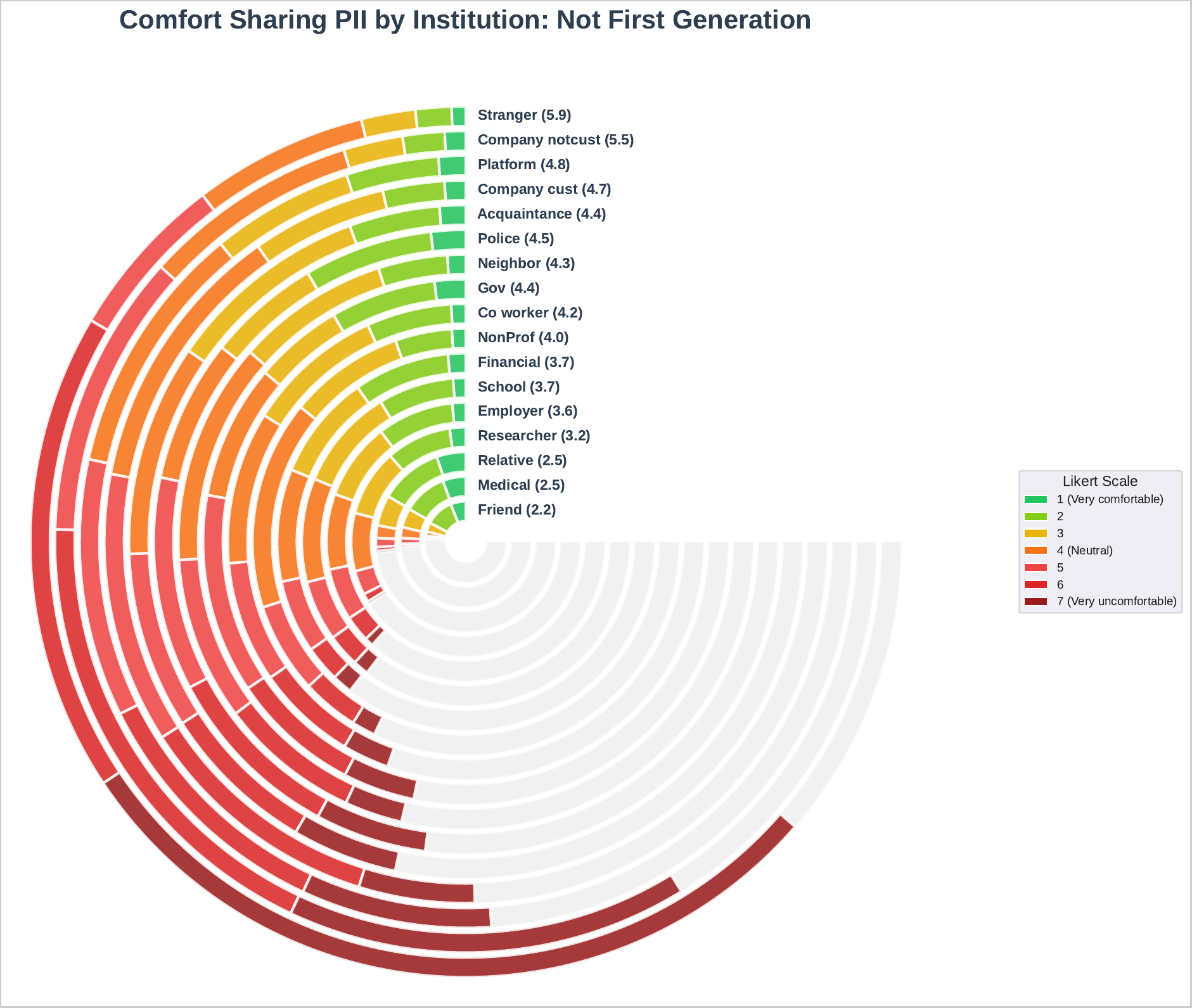}
    \caption{Comfort Sharing PII by Institution and Demographics First Generation. Arc length represents mean discomfort level, ranging from 1 when very comfortable (green) to 7 when very uncomfortable (red). Colored segments show the full breakdown of Likert responses.}
    \Description{Circular stacked bar chart showing comfort levels for sharing personal information with 17 institutions by first-generation status. Each institution is represented by a horizontal bar bent into an arc, divided into colored segments for the 7-point Likert scale. Green segments indicate comfort, yellow/orange neutral responses, and red segments discomfort.}
    \label{fig:arcdemplotsfirtgen}
\end{figure*}

\begin{figure*}
    \centering
    \includegraphics[width=.3\linewidth]{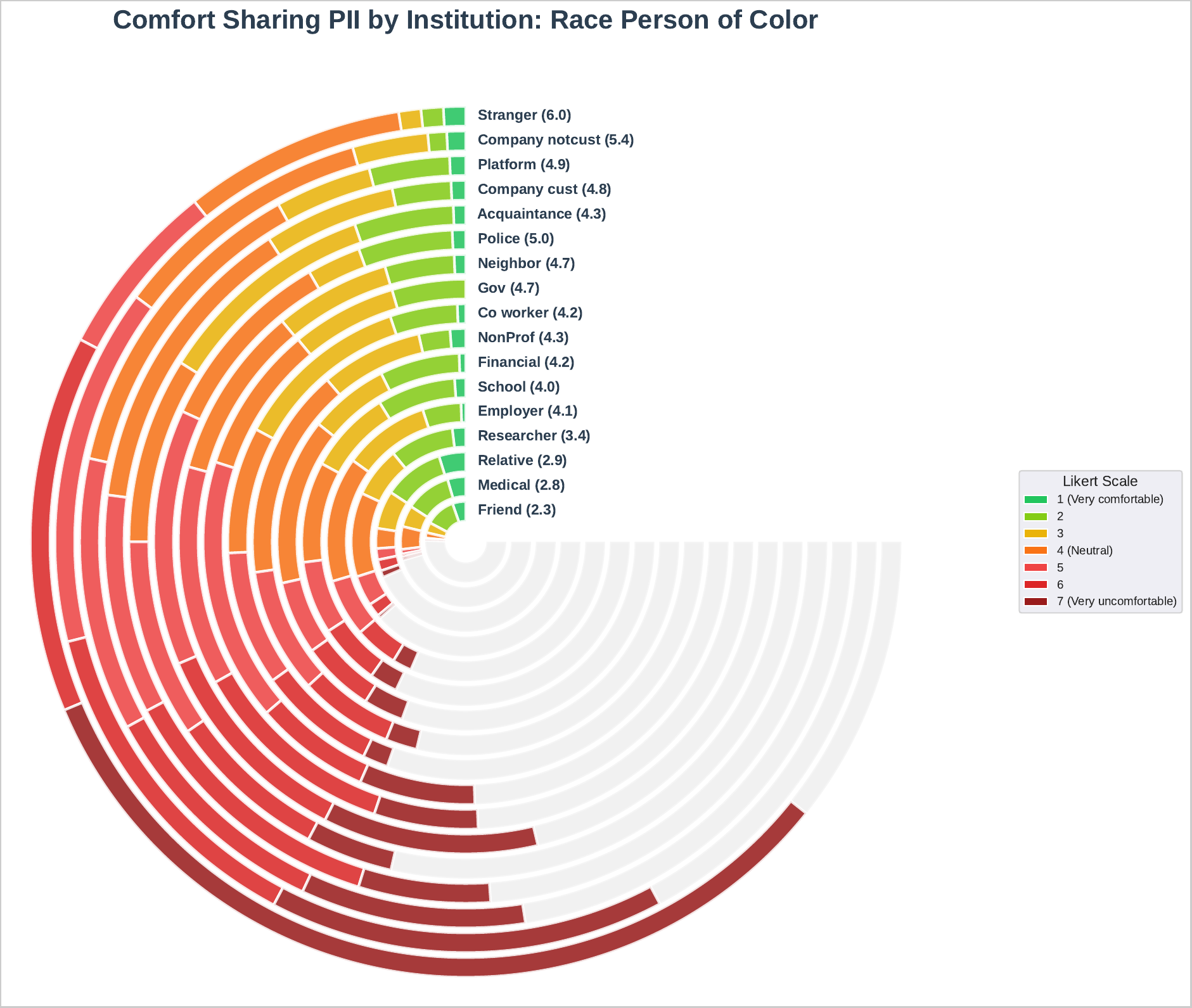}
    \includegraphics[width=.3\linewidth]{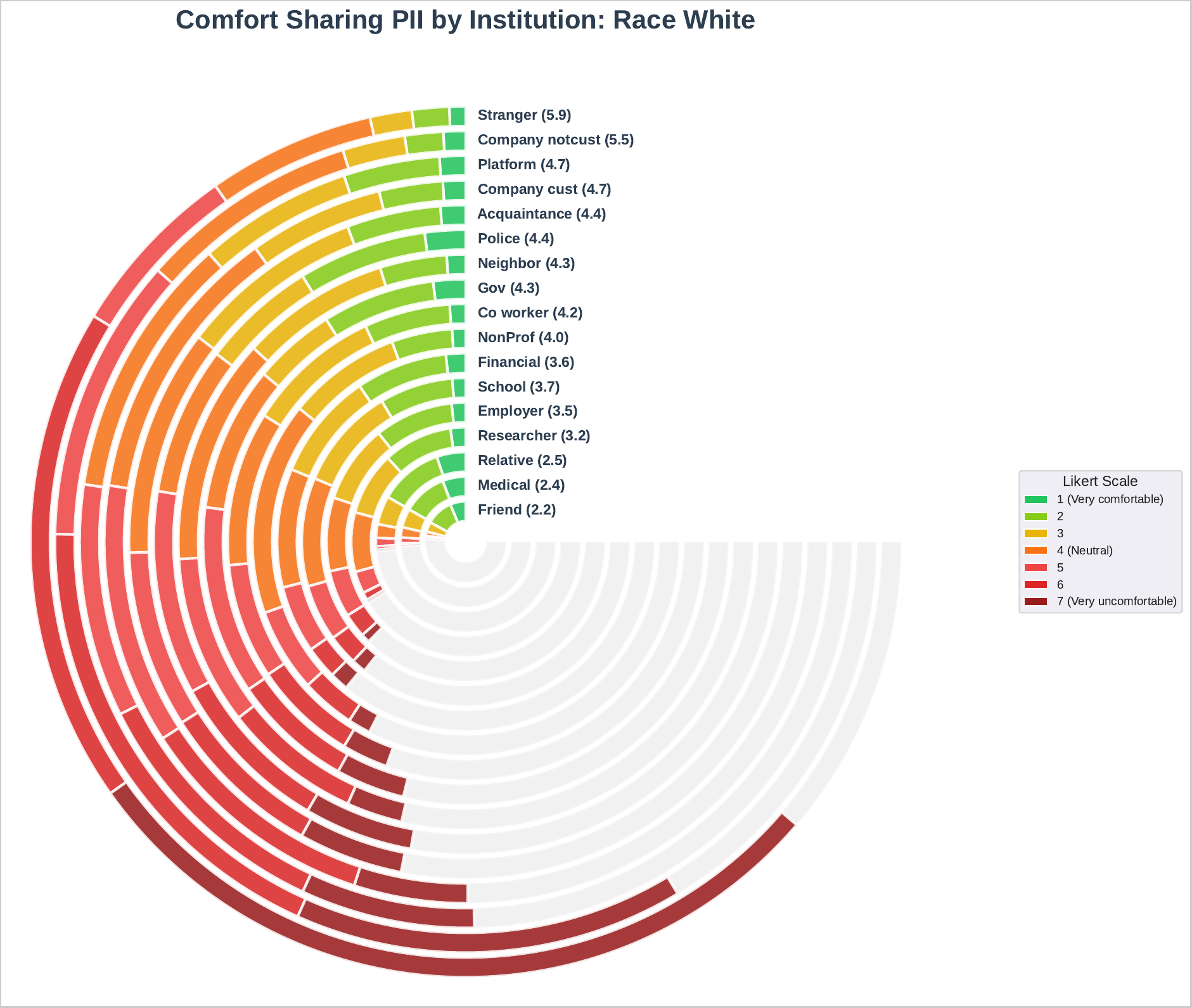}
    \includegraphics[width=.3\linewidth]{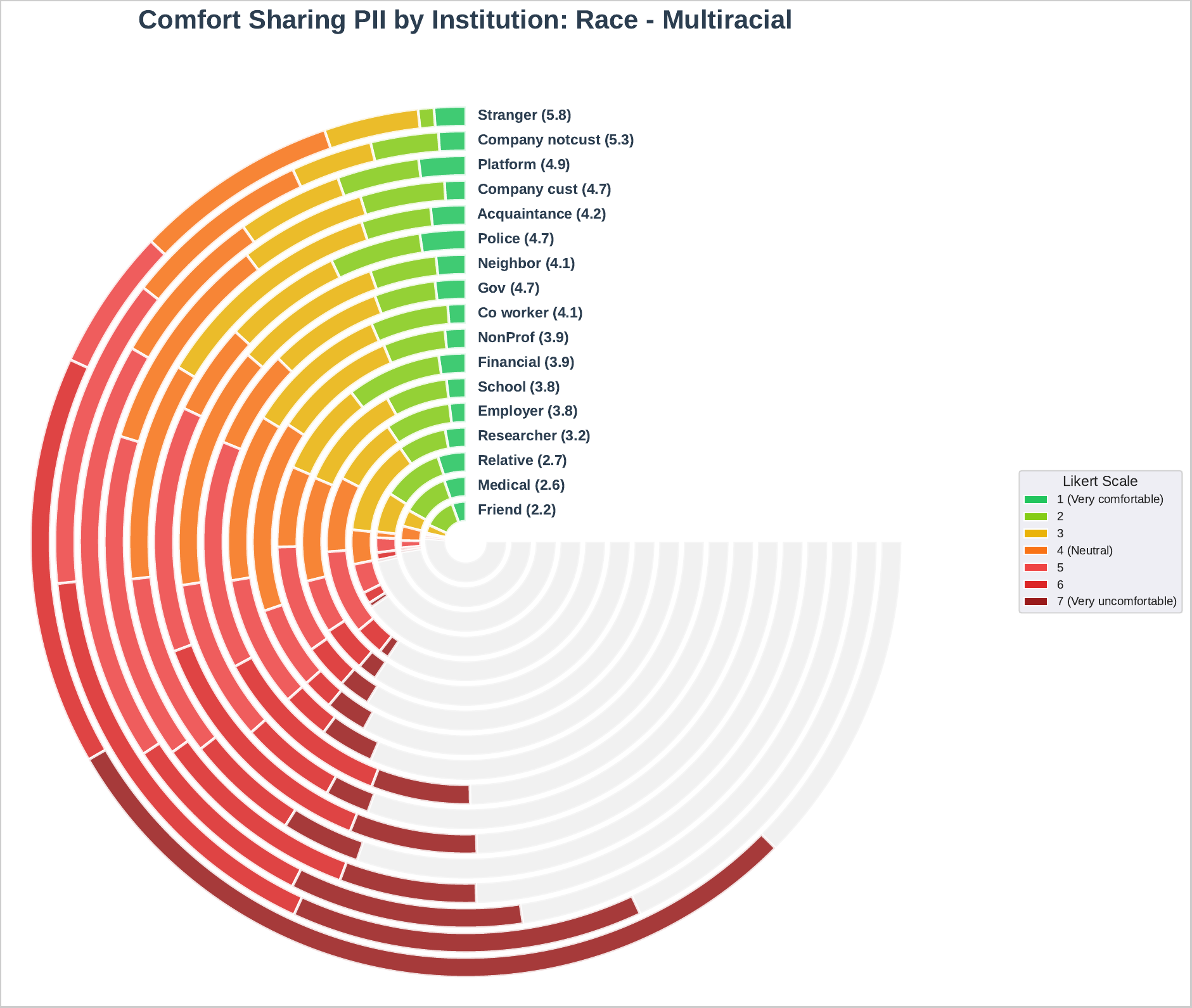}
    \caption{Comfort Sharing PII by Institution and Race. Arc length represents mean discomfort level, ranging from 1 when very comfortable (green) to 7 when very uncomfortable (red). Colored segments show the full breakdown of Likert responses.}
    \Description{Circular stacked bar chart showing comfort levels for sharing personal information with 17 institutions by race. Each institution is represented by a horizontal bar bent into an arc, divided into colored segments for the 7-point Likert scale. Green segments indicate comfort, yellow/orange neutral responses, and red segments discomfort.}
    \label{fig:arcdemplotsrace}
\end{figure*}

\begin{figure*}
    \centering
    \includegraphics[width=.3\linewidth]{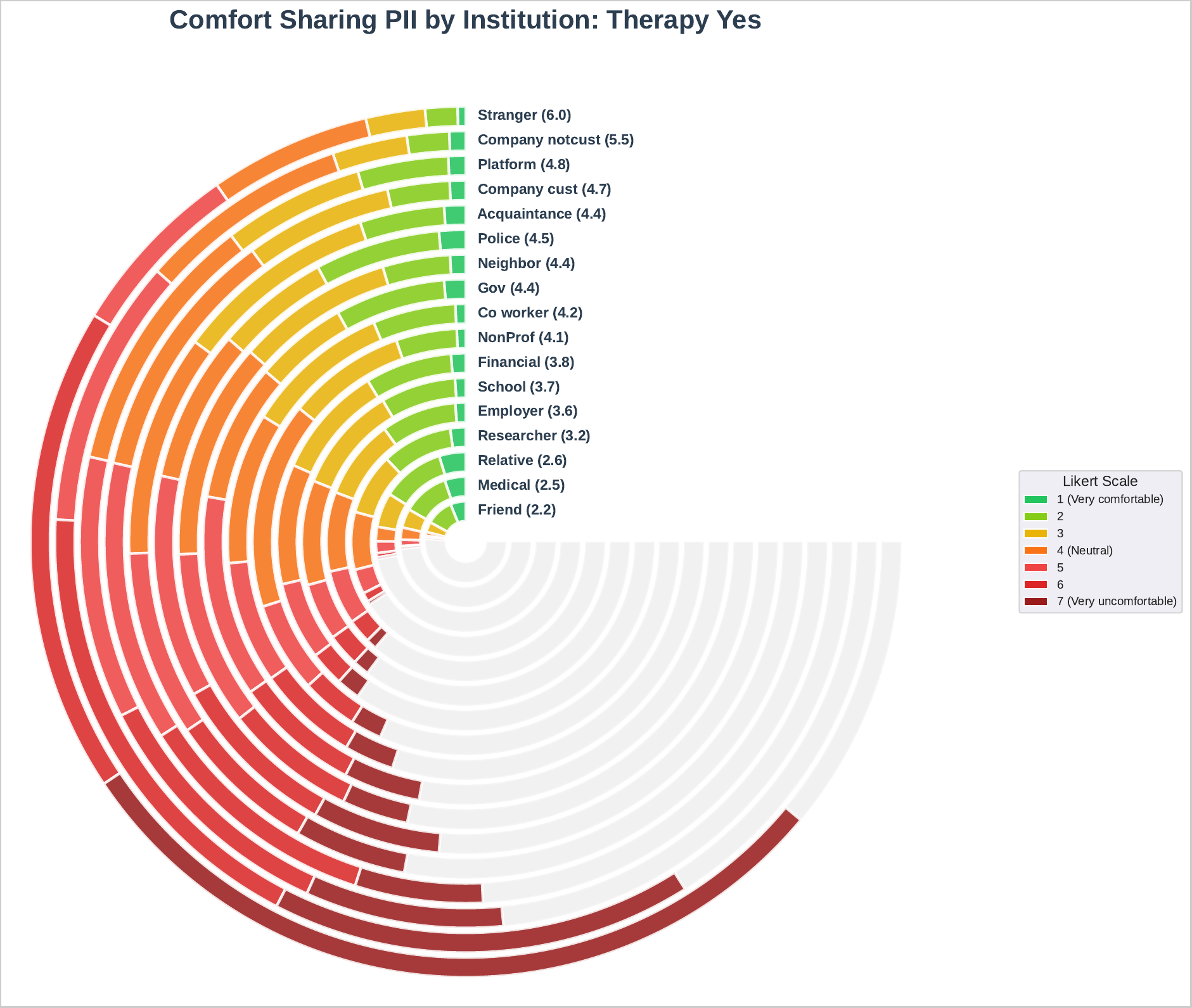}
    \includegraphics[width=.3\linewidth]{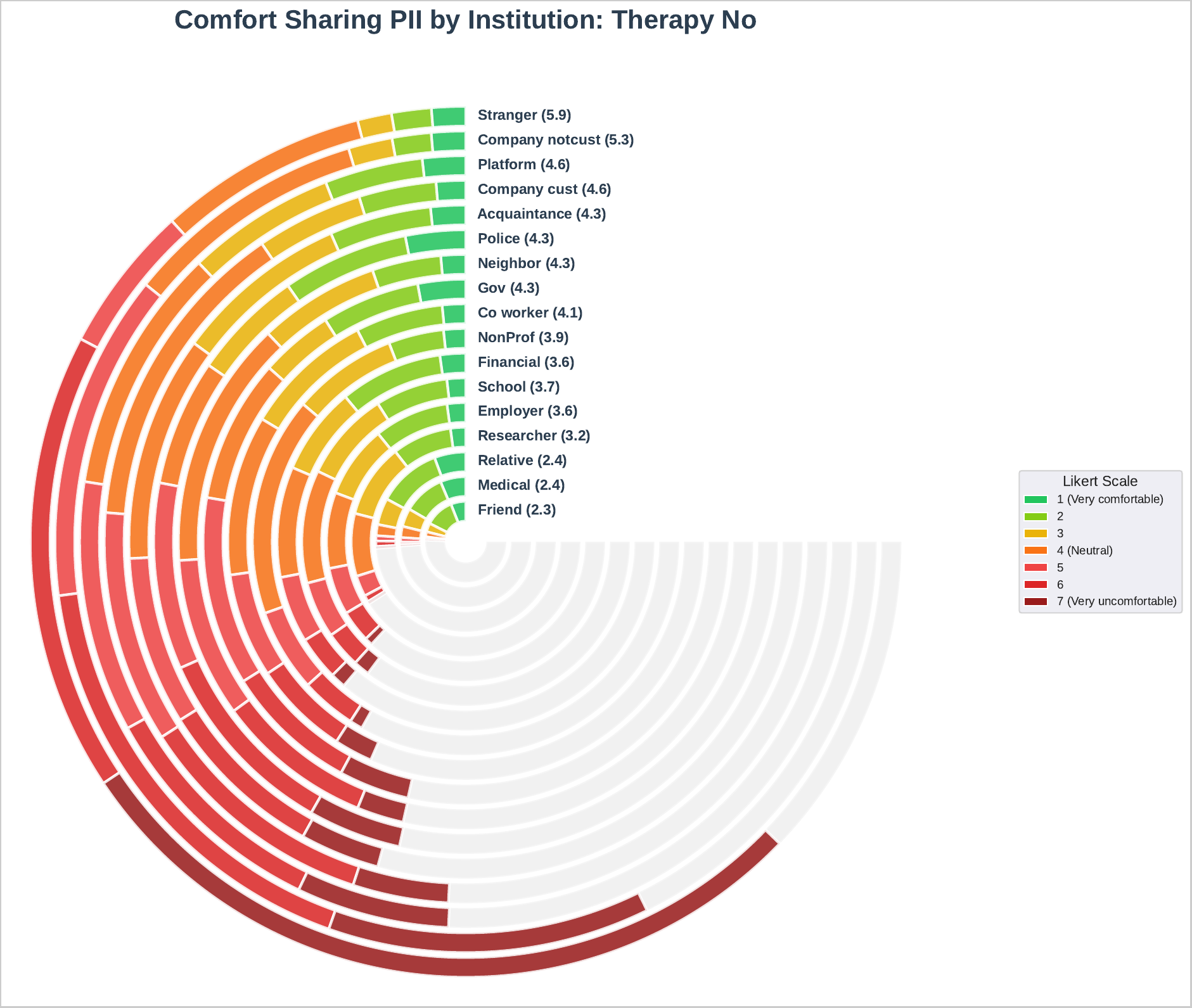}
    \caption{Comfort Sharing PII by Institution and Therapy Experience. Arc length represents mean discomfort level, ranging from 1 when very comfortable (green) to 7 when very uncomfortable (red). Colored segments show the full breakdown of Likert responses.}
    \Description{Circular stacked bar chart showing comfort levels for sharing personal information with 17 institutions by therapy experience. Each institution is represented by a horizontal bar bent into an arc, divided into colored segments for the 7-point Likert scale. Green segments indicate comfort, yellow/orange neutral responses, and red segments discomfort.}
    \label{fig:arcdemplotstherapy}
\end{figure*}

\begin{figure*}
    \centering
    \includegraphics[width=.3\linewidth]{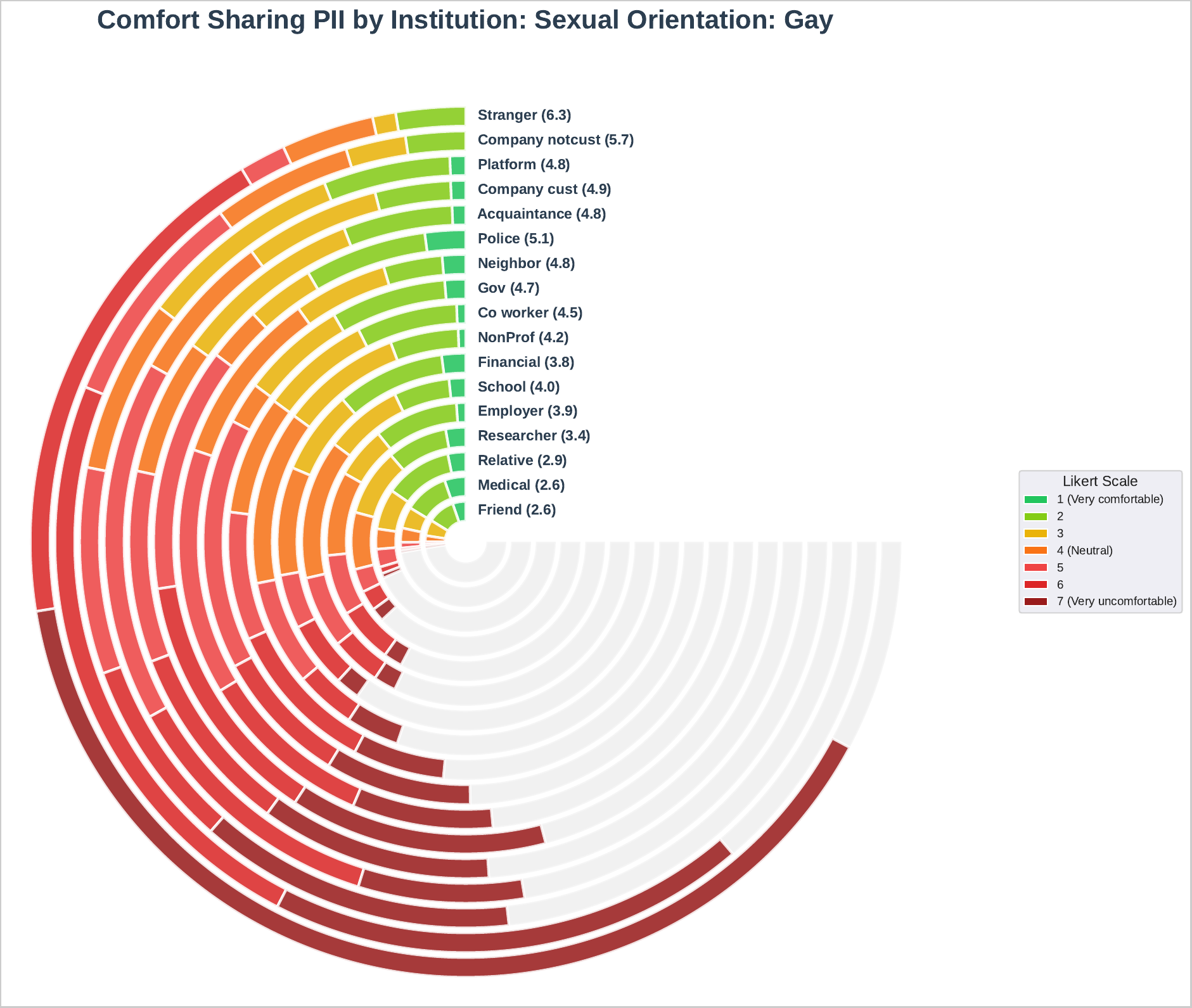}
    \includegraphics[width=.3\linewidth]{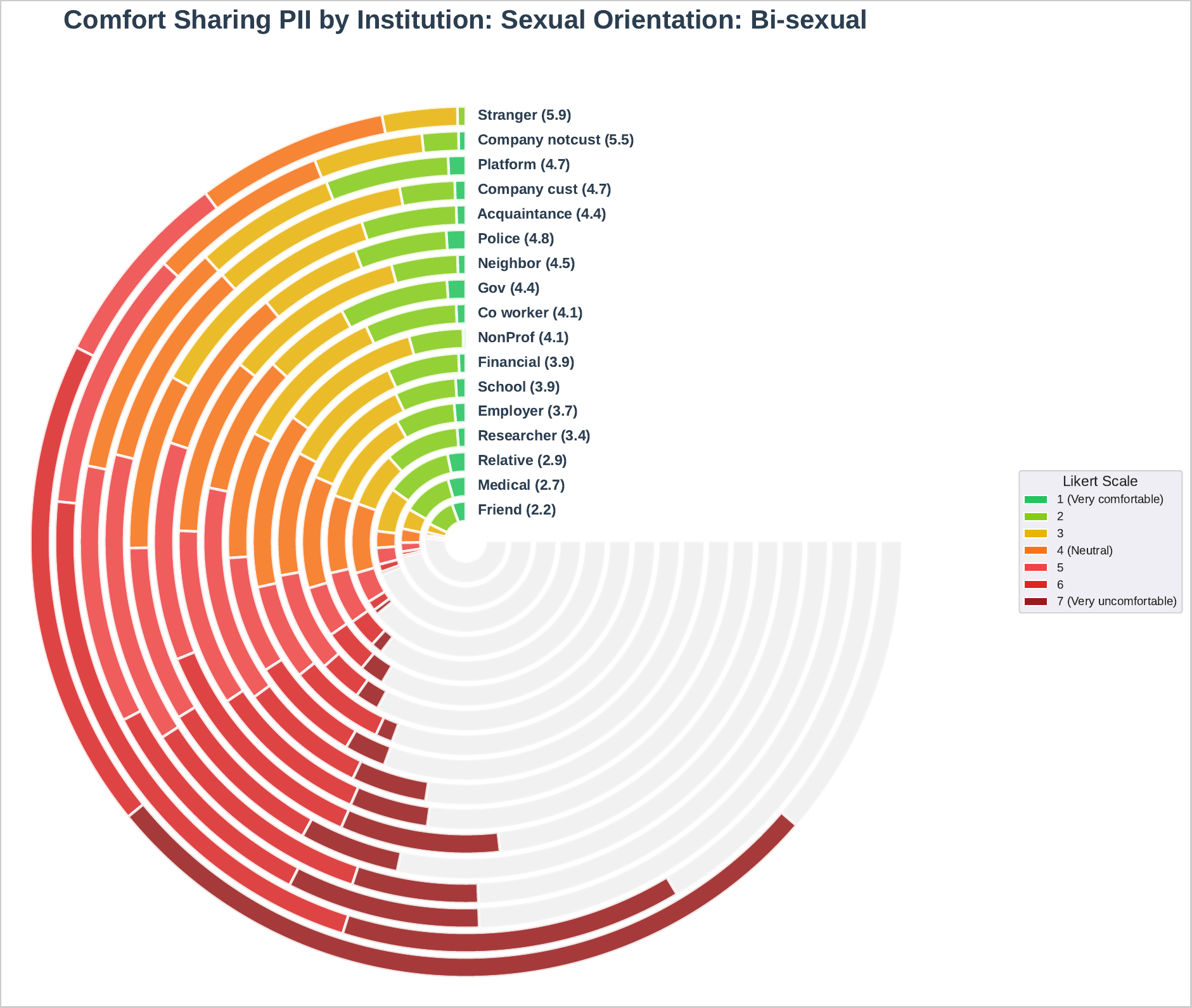}
    \includegraphics[width=.3\linewidth]{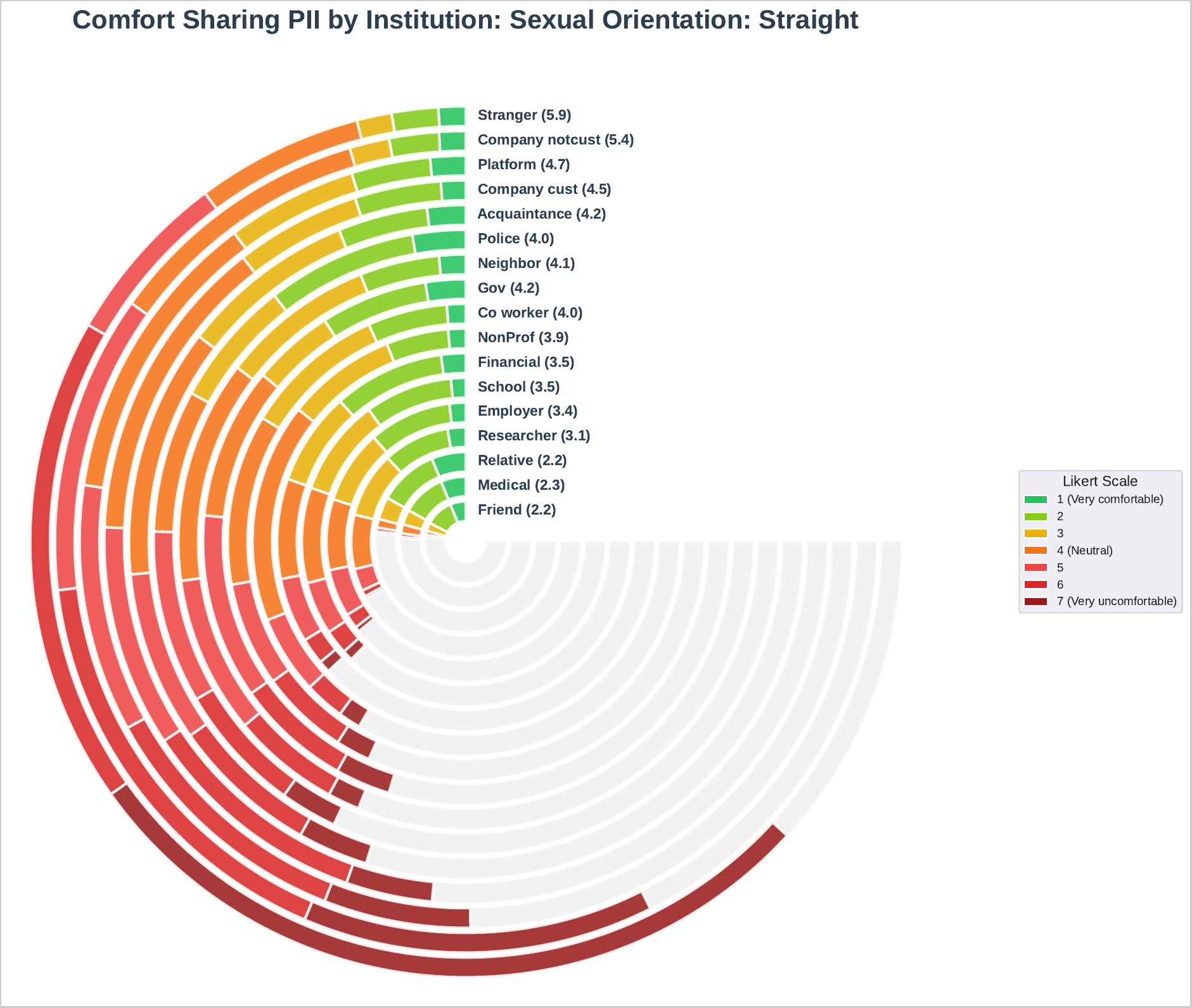}
    \includegraphics[width=.3\linewidth]{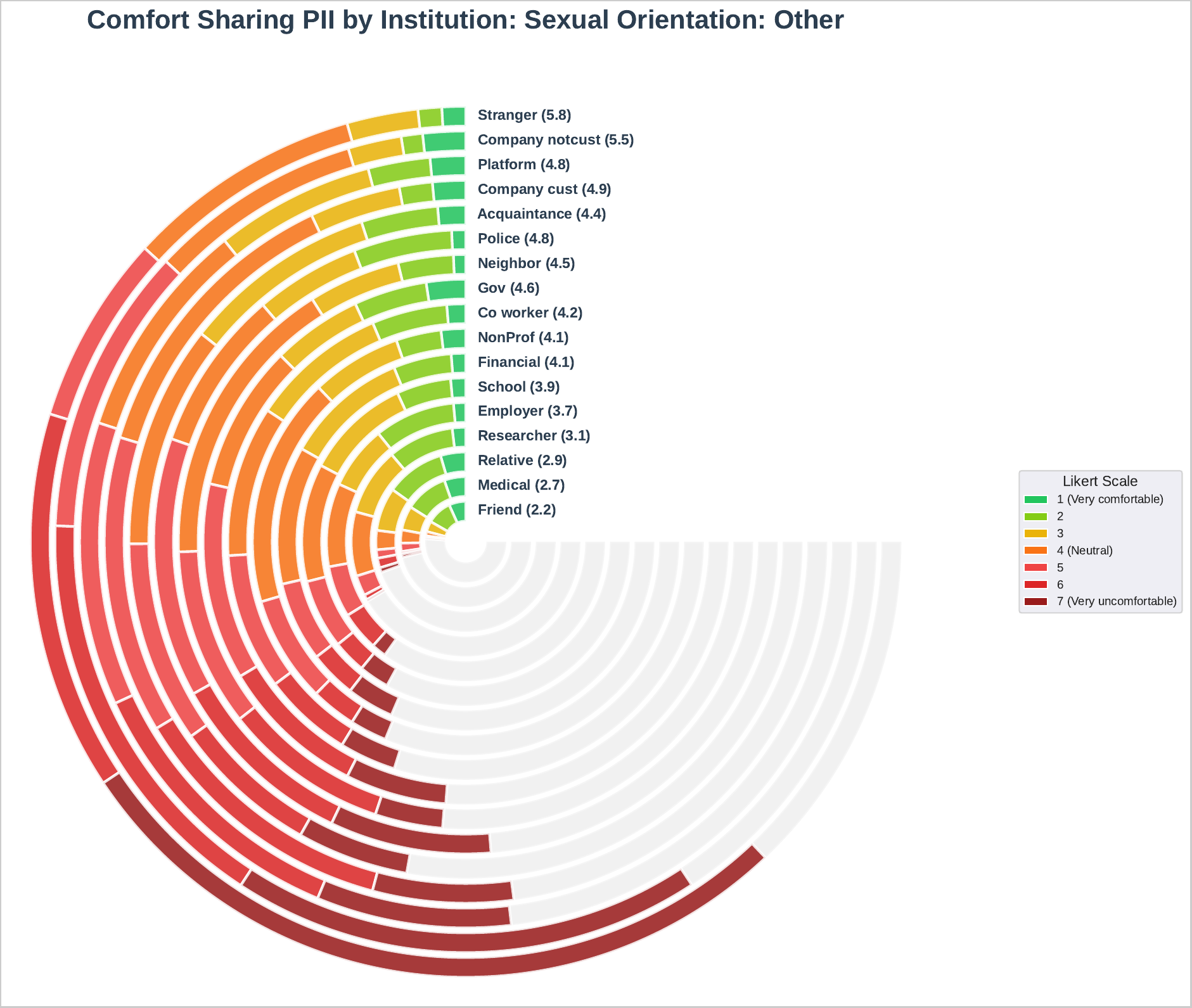}
    \caption{Comfort Sharing PII by Institution and Sexual Orientation. Arc length represents mean discomfort level, ranging from 1 when very comfortable (green) to 7 when very uncomfortable (red). Colored segments show the full breakdown of Likert responses.}
    \Description{Circular stacked bar chart showing comfort levels for sharing personal information with 17 institutions by sexual orientation. Each institution is represented by a horizontal bar bent into an arc, divided into colored segments for the 7-point Likert scale. Green segments indicate comfort, yellow/orange neutral responses, and red segments discomfort.}
    \label{fig:arcdemplotsSO}
\end{figure*}

\end{document}